\documentclass[peerreview,draftcls,onecolumn,11pt,a4paper]{IEEEtran}

\usepackage{amsmath,amsfonts,amsbsy,amssymb}
\usepackage{algorithmic}
\usepackage{array}
\usepackage[caption=false,font=normalsize,labelfont=sf,textfont=sf]{subfig}
\usepackage{textcomp}
\usepackage{stfloats}
\usepackage{url}
\usepackage{verbatim}
\usepackage{color}
\usepackage{graphicx}
\usepackage{cite}
\usepackage{hyperref}
\usepackage{setspace}
\usepackage{euscript}

\usepackage[left=1.25in,right=1.25in,top=0.9in,bottom=0.9in]{geometry}

\let\OLDthebibliography\thebibliography
\renewcommand\thebibliography[1]{
  \OLDthebibliography{#1}
  \setlength{\parskip}{-1.3pt}
  \setlength{\itemsep}{-1.3pt plus 0.3ex}
}

\hypersetup{
	colorlinks=true,
	linkcolor=blue,
	citecolor=blue,
	filecolor=magenta,      
	urlcolor=cyan,
}

\def\BibTeX{{\rm B\kern-.05em{\sc i\kern-.025em b}\kern-.08em
    T\kern-.1667em\lower.7ex\hbox{E}\kern-.125emX}}
\usepackage{balance}

\newcommand{\bm}[1]{{\mathbf{#1}}}
\newcommand{\Es}{{\mathbb{E}}}          
\newcommand{\Prob}{{\mathbb{P}}}        

\newcommand{\diag}{{\text{diag}}}

\newcommand{\x}{\bm x}

\newcommand{\vb}{\bm v}
\newcommand{\s}{\bm s}
\newcommand{\y}{\bm y}
\newcommand{\w}{\bm w}
\newcommand{\bu}{\bm u}
\newcommand{\F}{\bm F}
\newcommand{\G}{\bm G}
\newcommand{\Gammab}{\boldsymbol{\Gamma}}
\newcommand{\gammab}{\boldsymbol{\gamma}}
\newcommand{\bb}{\bm b}
\newcommand{\cb}{\bm c}
\newcommand{\pb}{\bm p}
\newcommand{\bk}{\bm k}
\newcommand{\ab}{\bm a}
\newcommand{\herm}{\text{H}}
\newcommand{\trasp}{\text{T}}
\newcommand{\eqdef}{\triangleq}

\newcommand{\pot}{\EuScript{P}}
\newcommand{\gain}{\EuScript{G}}
\newcommand{\rate}{\EuScript{R}}
\newcommand{\cost}{\EuScript{C}}

\def\bdm#1\edm{\begin{displaymath}#1\end{displaymath}}
\def\be#1\ee{\begin{equation}#1\end{equation}}
\def\barr#1\earr{\begin{align}#1\end{align}}

\begin{document}
\title{\setlength{\baselineskip}{10.0mm}
Integrating sensing and communications: 
Simultaneously transmitting and reflecting digital coding metasurfaces}
\author{Francesco Verde, Vincenzo Galdi, Lei Zhang, and Tie Jun Cui}

\maketitle

\begin{abstract}
\setlength{\baselineskip}{5.0mm}
Wireless networks are undergoing a transformative shift, driven by the crucial factors of cost effectiveness and sustainability. 
Digital coding metasurfaces (DCMs) might play a key role in realizing cost-effective digital modulators by harnessing  energy embedded in electromagnetic waves traversing through the air.
Integrated sensing and communication (ISAC) optimize power and spectral resources by combining sensing and communication functionalities on a shared hardware platform.
This article presents a tutorial-style overview of the applications and advantages of DCMs in ISAC-based networks.
Emphasis is placed on the dual-functionality of ISAC, necessitating the design of DCMs with simultaneously transmitting and reflecting (STAR) capabilities for comprehensive space control. 
Additionally, the article explores key signal processing challenges and outlines future research directions stemming from the convergence of ISAC and emerging STAR-DCM technologies.

\end{abstract}

\section{Introduction}

The historical linear model of technological progress is now transitioning towards a circular economy, emphasizing principles of recycling and sharing for sustainable growth and resource conservation.
The integration of information and communication technologies (ICT) is seen as a key enabler for optimizing policies during this transition. The concepts of recycling and sharing in wireless communications have roots dating back to the late 1940s and early 1960s, with Stockman's ``reflected-power communications'' \cite{Stock.1948} and Mealey's ``integrated sensing and communications'' (ISAC) \cite{Mealey.1963}. These once-theoretical concepts have now become tangible realities, thanks to advancements in signal processing algorithms and electronic technologies.

Reflected-power communications seamlessly integrate recycling by eliminating the need for active analog components to generate a carrier wave. Utilizing dedicated or naturally available sources, such as digital TV broadcasting and cellular systems, these systems inherently harvest energy for minimal power consumption and cost-effective implementation.
Traditional approaches, like radio-frequency identification and reflectarray antennas, are being complemented by emerging technologies such as electromagnetic (EM) ``metamaterials'' or ``metasurfaces,'' with deeply subwavelength elements, which exhibit unique properties, enabling advanced field manipulations for various applications \cite{Capolino}.
In wireless communications, metasurfaces are typically referred to as {\em reconfigurable intelligent surfaces (RISs)} \cite{DiRenzo.2022,Nerini.2024}, since the EM properties
of their constitutive elements need to be reconfigured over time as a function of the time-varying propagation and network conditions. 
RISs can be characterized by continuous macroscopic medium parameters or described in a digital manner.
{\em Digital coding metasurfaces} (DCMs) \cite{Cui:2014dm} are digital RISs, whose physical parameters, such as amplitude, phase, and polarization, are represented in a digital format using binary codes and controlled/programmed via electronic circuits such as field-programmable gate arrays (FPGAs).
For instance, the elements constituting 1-bit DCMs take on the opposite phase values $0$ and $\pi$ radians that mimic 
the digits ``0'' and ``1'', respectively. The concept of DCMs can be extended from $1$-bit to multi-bits by discretizing 
the phase over the whole round angle.
This allows real-time manipulation of continuous-time signals through binary coding, marking a successful convergence of physics and digital signal processing over the past decade.

ISAC embodies resource sharing by utilizing a single hardware setup for both sensing and communication functions. This integration enhances spectral efficiency and reduces hardware expenses. Through joint design, ISAC enables a flexible trade-off across various domains, including time, frequency, code, and spatial aspects. 
This article focuses on an ISAC model grounded in concurrently transmitting information while extracting information from scattered echoes \cite{YCui.2021}. It builds on previous research covering various aspects of ISAC, including waveform options, architecture, resource allocation, antenna deployment, and cooperative/non-cooperative coordinated operations.
The central question is how to seamlessly integrate recycling (DCMs) and sharing (ISAC) to create advanced technology entities for value-added ICT services. 
Such an integration mainly depends on the operational mode of the DCM.

\begin{figure}
	\centering
	\includegraphics[width=1\linewidth]{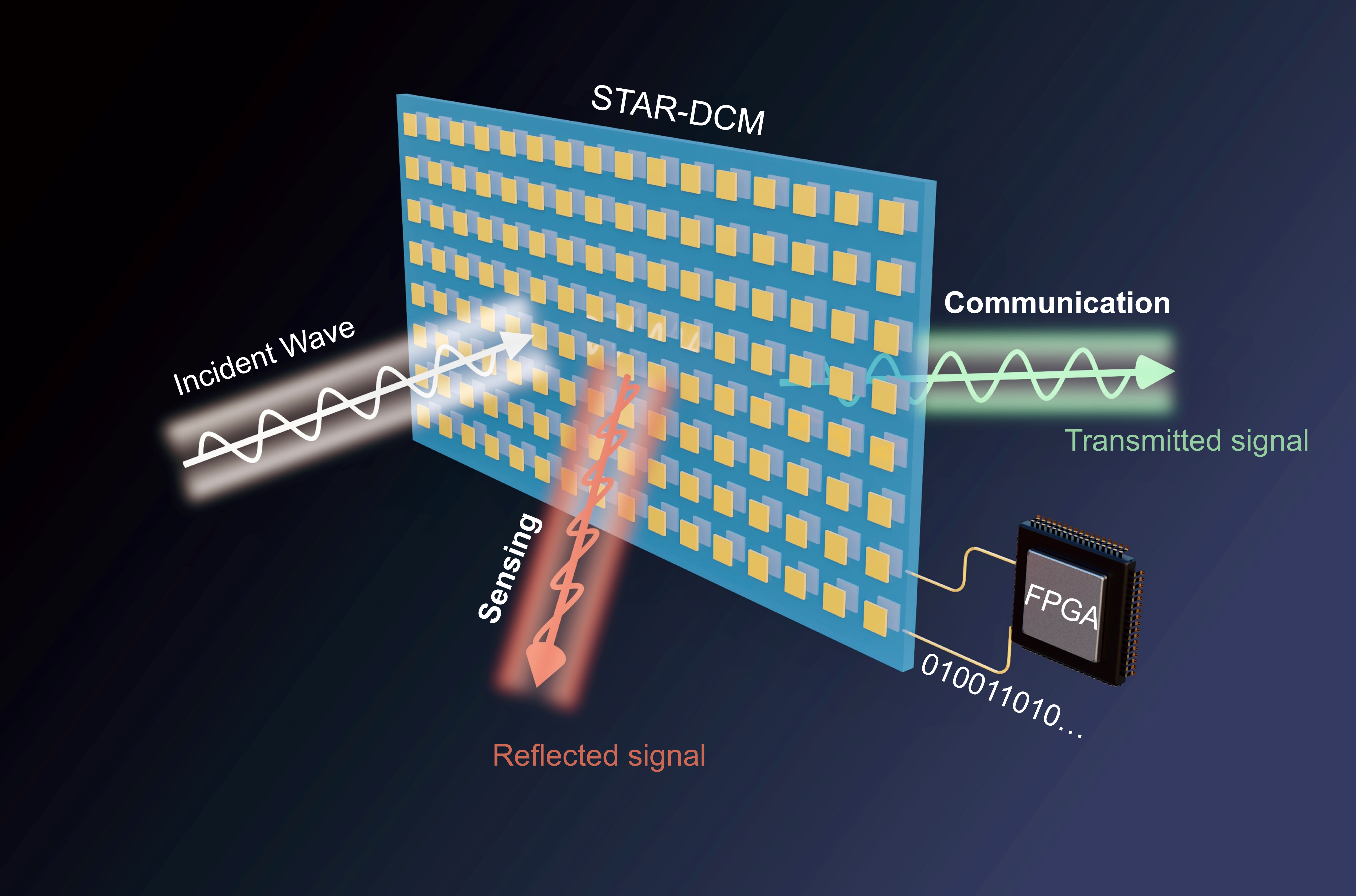}
	\caption{Conceptual illustration of ISAC by means of a STAR-DCM.}
	\label{fig:figuregeometry}
\end{figure}

Traditional DCMs can work in either transmitting  or reflecting mode, thus
controlling EM waves only on one side of their interface.
Many existing works on DCM-enabled ISAC systems
consider the case where DCMs are only able to reflect the incident wireless
signals \cite{Esma.2022,Esma_Arxiv.2024}, which will be referred to as 
reflecting-only DCMs. In this case, all the devices involved in 
the ISAC process must be located on the same
side of the DCM, thus leading to half-space coverage.
Recent developments in {\em simultaneously transmitting and reflecting} (STAR) DCMs \cite{Zhang:2018tr,Bao:2021pr,Mu.2022},
also referred to as {\em intelligent omni-surfaces} \cite{Xu.2022}, 
enable  full-space control of EM fields. This forms the basis for more effective ISAC networks, where signals can be reflected for state 
measurements or refracted (i.e., transmitted) for information conversion, 
as conceptually illustrated in Fig. \ref{fig:figuregeometry}.
In this case, compared to reflecting-only DCMs, 
one can develop highly flexible full-space DCM-enabled ISAC systems,
allowing communication and sensing devices to be be located on 
both sides of the DCM \cite{Wang.2023,Liu.2023}, thereby broadening 
the application potential for ISAC.

This article provides a signal processing-oriented tutorial on STAR-DCMs, exploring their 
fundamental properties, assessing ISAC architectures with STAR-DCMs, 
and discussing EM design and signal processing for communication and sensing. 
Moreover, experimental results from prototype fabrication highlight the 
promising potential of implementing ISAC networks via STAR-DCMs.

\section{Historical perspective}
\label{sec:history}

This section outlines the historical background of ISAC and STAR-DCMs. First, 
we highlight the pioneering contributions and the past key developments of the two technologies separately,
and, then, we examine their recent integration.

\subsection{Co-existence between sensing and communication technologies}

Sensing and communication functionalities can be mixed in different ways. The {\em coexistence} approach treats them as separate end-goals, operating together to avoid conflicts and maximize performance. However, it consumes resources and limits spatio-temporal degrees of freedom.
Moving beyond coexistence, the fundamental principle of ISAC is to cohesively design sensing and communication operations, allowing the sharing of hardware, power, and bandwidth resources.
Sensing applications cover ultra-high accuracy localization, tracking, imaging, mapping, environmental reconstruction, augmented human sensing, and gesture/activity recognition.
ISAC systems trace back six decades to the introduction of dual-function radar communication (DFRC) \cite{Mealey.1963}. DFRC systems, historically radar-centric, integrated communication as a secondary function within existing radar infrastructures. In the late 20th century, a communication-centric approach emerged with packet-switching technologies. Recent ISAC networks incorporate positioning techniques within existing communication systems, evolving into a sophisticated paradigm \cite{YCui.2021}, which facilitates diverse interactions between sensing and communication. 
Additional sensors beyond radar and a broader range of sensing services beyond localization are involved in \cite{YCui.2021}.

\subsection{Independent control of phase responses in reflection and transmission modes}

Reflected-power communications, originating from Stockman's 1948 paper \cite{Stock.1948}, involve modulating an incident EM wave while reflecting it. 
Metasurfaces can be considered a natural evolution of this concept.
Traditionally, metasurfaces have efficiently operated in either reflection or transmission modes. In 2017, a groundbreaking STAR metasurface demonstrated distinct polarization-dependent transmission and reflection properties \cite{Cai:2017he}. 
Subsequently, researchers have explored metasurfaces controlling both reflection and transmission responses at distinct frequencies in various studies \cite{Wang:2018sr, Yue:2019da, Liu:2020rs}. In 2019, a programmable STAR-DCM was introduced, incorporating a reflection-transmission amplitude code and an independent phase code \cite{Wu:2019dm}. 
Several programmable STAR-DCM platforms now enable real-time control over transmission, reflection, duplex, and absorption responses \cite{Wang:2021ar, Zhang:2022io, Hu:2022ai, Wang:2022bh, Yin:2023rt}. Recent research has explored STAR-DCMs for comprehensive manipulations of circularly polarized waves in full space \cite{Sun:2023tr} and high-efficiency polarization conversion \cite{Qin:2023tr}.

\subsection{STAR-DCM-aided ISAC systems}
The integration of STAR-DCMs into ISAC networks has a very recent history. 
The proposal of enabling ISAC by STAR-DCMs has been originaly made in 
\cite{Wang.2023,Liu.2023}. In such works, the whole space 
is divided into two half-spaces, namely the sensing space and the
communication space. The signal from the transmitter is 
split at the STAR-DCM to sense targets in the sensing
space and deliver data to users in the communication space.
Joint optimization of transmit beamforming and phase
shift of a STAR-DCM has been carried out in \cite{Eghbali.2024}.
The potentials of STAR-DCMs have been recently exploited in vehicular 
scenarios \cite{Meng.2024,Li.2024}. In these use cases, 
a STAR-DCM is equipped on the outside surface of a vehicle
to improve the communication service of the in-vehicle user equipment 
and, simultaneously, track and sense the vehicle with the help of nearby 
roadside units.
On one hand, the merging of STAR-DCMs and ISAC allows
to extend network coverage. On the other hand, it inevitably exacerbates the
competition for the limited bandwidth resources in wireless systems, which may impair
ISAC performance. More recently, to balance bandwidth demand and 
ISAC performance, non-orthogonal multiple access (NOMA)
has been suggested as 
a promising approach \cite{Sun.2024,Wang.2024,Xue_2024,Wei.2024}. 

\section{Electromagnetic design and implementation of STAR-DCMs}
\label{sec:EMdesign_implementation}

This section addresses the design and practical implementation of STAR metasurfaces in hardware, essential for potential ISAC applications. It offers insights into various architectures where successful prototypes demonstrate feasibility.

Conventional reflection-only DCMs typically feature a single-layer structure with a metal-backed dielectric substrate and printed metallic elements. The local phase response is manipulated by adjusting element size or, in programmable setups, reconfiguring responses with electronic switches like diodes \cite{Cui:2014dm}.
However, single-layer, non-magnetic transmission-type metasurfaces face limitations in efficiency and phase response control. Therefore, multilayered configurations are often adopted to operate in both transmission and reflection modes. Yet, achieving simultaneous and independent control of these responses presents a challenge due to their inherent interconnection.

In \cite{Cai:2017he}, a {\em polarization-division} strategy was proposed, with a metasurface element featuring four metallic layers and three dielectric spacers that selectively allows the transmission of one linear polarization. This design achieves independent and simultaneous control of reflection and transmission responses.
In \cite{Wang:2018sr},  a {\em frequency-division} strategy  was put forward, with a metasurface element composed of patterned and perforated layers separated by a dielectric substrate, which achieves independent control of transmission and reflection responses at two design frequencies.

Both designs above have fixed functionalities, relying on locally tuned dimensions. In what follows, we illustrate four examples of STAR-DCMs based on reconfigurable elements.

\begin{figure*}
	\centering
	\includegraphics[width=1\linewidth]{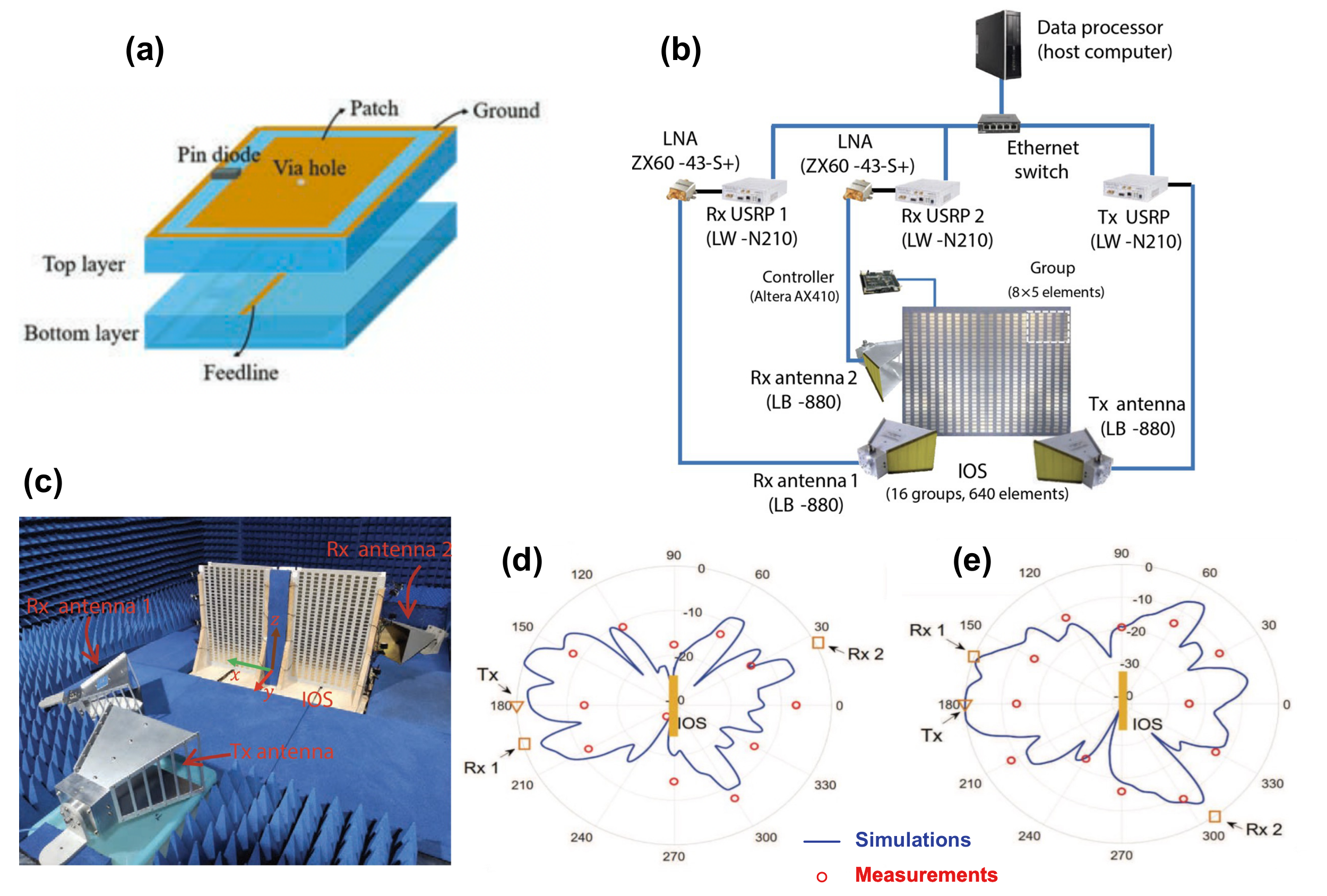}
	\caption{Energy-splitting STAR-DCM. (a) Schematic of metasurface element. (b) Conceptual illustration and wireless communication prototype. (c) Photograph of experimental setup for verifying the beamforming on both side of the metasurface. (d), (e) Representative (simulated and measured) scattering  patterns. [Used from \cite{Zhang:2022io}, \copyright 2022 IEEE]}
	\label{fig:HW1}
\end{figure*}

Figure~\ref{fig:HW1} illustrates an {\em energy splitting} STAR metasurface operating at 3.6 GHz \cite{Zhang:2022io}. The metasurface element, shown in Fig. \ref{fig:HW1}(a) comprises two mirror-symmetric layers, and utilizes two positive-intrinsic-negative (PIN) diodes for controlling transmission and reflection coefficients
up to four states. Though simple, this design has limitations in achieving independent control of transmission and reflection.
The prototype, shown in Fig.~\ref{fig:HW1}(b), consists of $640$ elements. Employed in a wireless communication prototype, the STAR metasurface demonstrates omni-beamforming capability. The experimental setup in Fig. \ref{fig:HW1}(c) features Tx and Rx antennas on either side of the metasurface, controlled by an FPGA through digital interfaces.
Figures \ref{fig:HW1}(d) and \ref{fig:HW1}(e) show the scattering patterns (i.e., the power scattered by the STAR metasurface as a function of the observation angle) for two representative configurations featuring a transmitter and two receivers at both sides of the metasurface.
These results demonstrate good agreement between simulations
(blue-solid curve) and measurements (red circles),
as well as the capability to attain directional beams for users on both sides of the STAR-DCM.

\begin{figure*}
	\centering
	\includegraphics[width=1\linewidth]{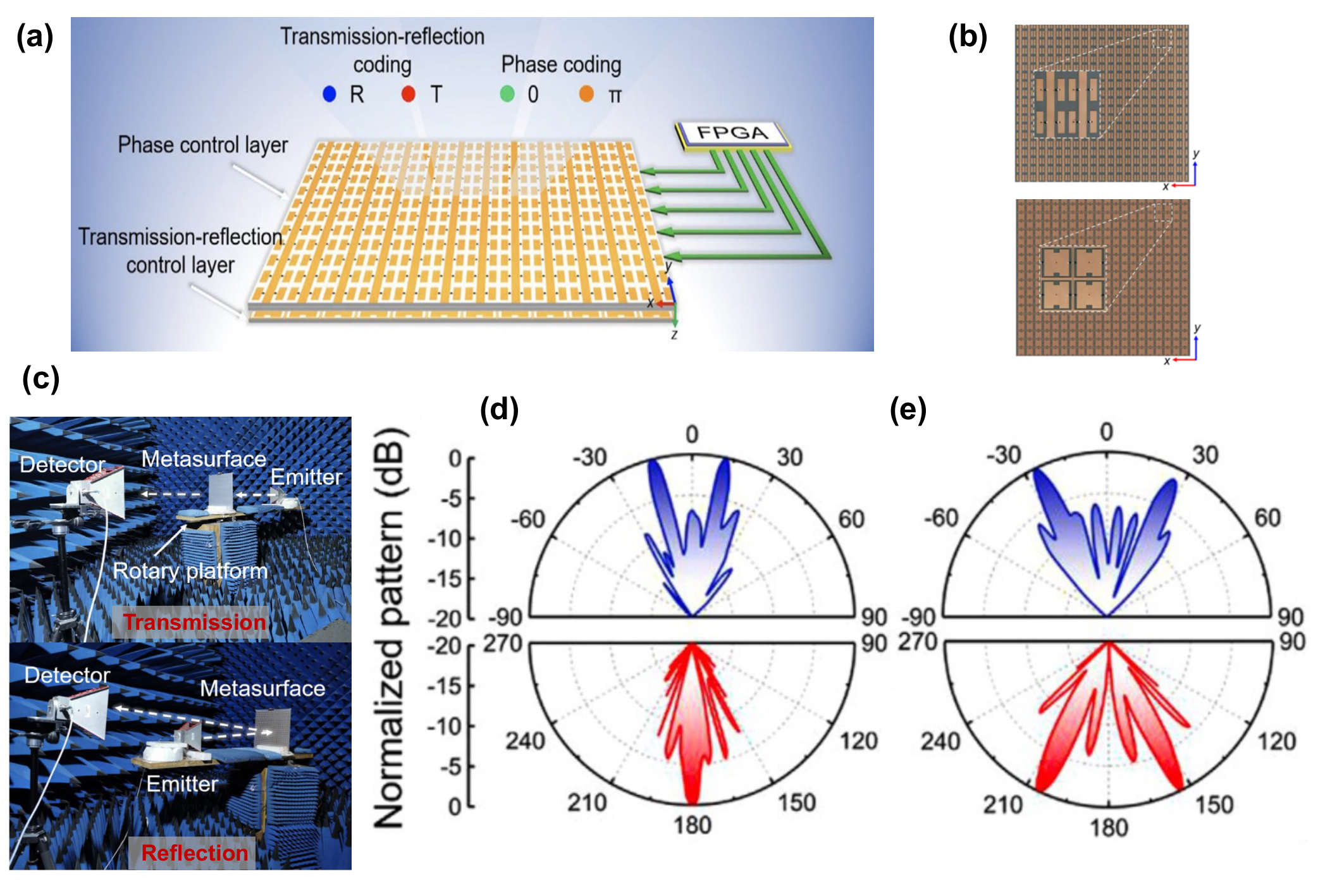}
	\caption{Mode-switching STAR-DCM. (a) Conceptual illustration. (b) Details of fabricated prototype, with metasurface element (for both layers) enlarged in the insets.
(c) Photographs of experimental setup for verifying dynamic beamforming in both transmission and reflection space.
 (d), (e) Representative measured scattering patterns, with blue and red color-coding indicating the two sides of the STAR-DCM. [Used from \cite{Hu:2022ai} with permission, \copyright 2022 Wiley]}
	\label{fig:HW2}
\end{figure*}

Figure~\ref{fig:HW2} illustrates
 a  {\em mode switching} configuration 
 operating at 6.15 GHz, and offering independent control of reflection and transmission. Specifically, Fig.~\ref{fig:HW2}(a) illustrates a conceptual model featuring  $20\times 20$ elements
 \cite{Hu:2022ai}, and Fig.~\ref{fig:HW2}(a) shows some details of the fabricated prototype. The metasurface element, shown in the insets of Fig. \ref{fig:HW2}(b), comprises two distinct PIN-diode-based units.  These stacked units enable independent  phase modulation and transmission/reflection control through a judicious combination of layers with an appropriate air gap.
 Figure \ref{fig:HW2}(c) shows the experimental setup, for measurements in transmission and reflection. The scattering patterns in Figs.~\ref{fig:HW2}(d) and \ref{fig:HW2}(e) exemplify duplex mode operation, achieving different beamforming functions on the two sides of the STAR-DCM.

\begin{figure*}
	\centering
	\includegraphics[width=1\linewidth]{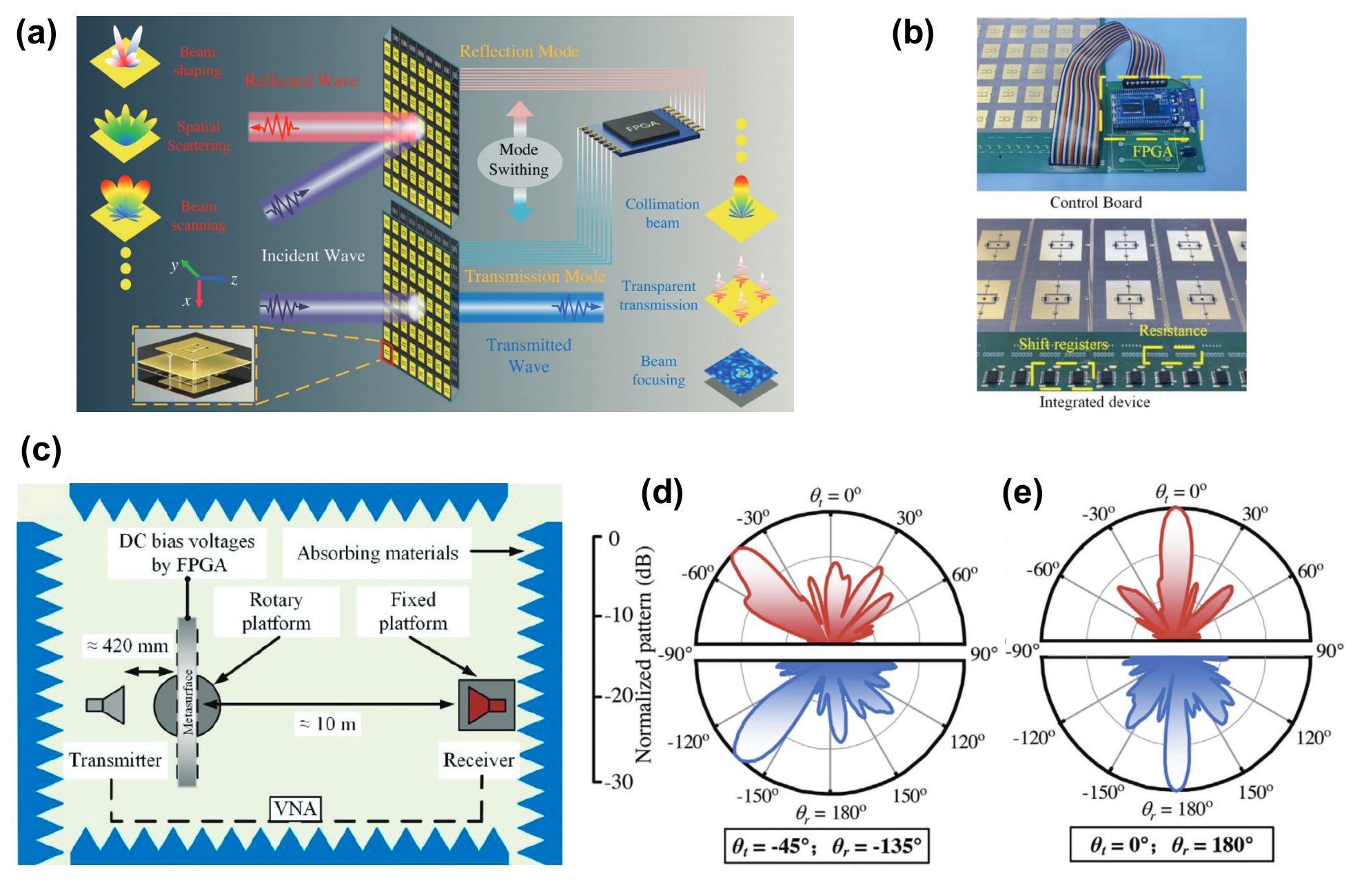}
	\caption{Alternative (low-profile) mode-switching STAR-DCM.
(a) Conceptual illustration, with metasurface element schematic shown in an inset.
(b) Details of fabricated prototype.
(c) Schematic of experimental setup.
(d), (e) Representative measured scattering patterns, with blue and red color-coding indicating the two sides of the STAR-DCM.
[Used from \cite{Yin:2023rt} with permission, \copyright 2024 Wiley]}
	\label{fig:HW3}
\end{figure*}

Figure~\ref{fig:HW3} presents a different implementation of the same principle operating at $3.5$ GHz\cite{Yin:2023rt}. 
Figure \ref{fig:HW3}(a) shows a conceptual illustration, with various possible operational modes.
The metasurface element, shown in the inset of Fig. \ref{fig:HW3}(a), utilizes a stack-up structure with two dielectric layers, four metallic layers and three PIN diodes, allowing switching between reflection and transmission modes with phase modulation. Figure \ref{fig:HW3}(b) shows some details of the
 fabricated STAR-DCM prototype, featuring $12\times12$ coding elements, whereas Fig. \ref{fig:HW3}(c) illustrates the experimental setup. As can be observed from the representative scattering patterns in Figs.~\ref{fig:HW3}(d) and \ref{fig:HW3}(e), this configuration has the
 capability for beam steering in full space.

\begin{figure*}
	\centering
	\includegraphics[width=1\linewidth]{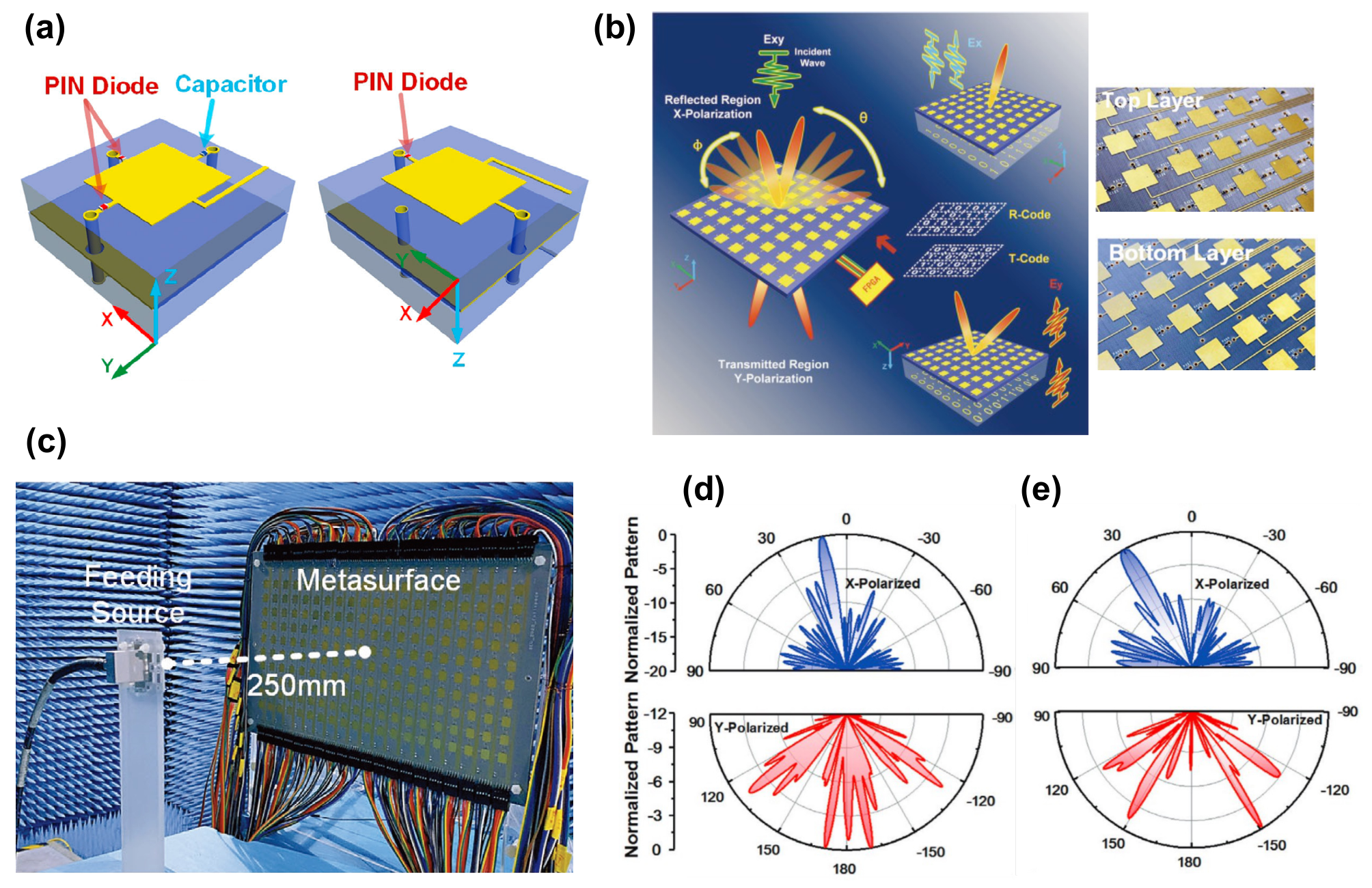}
	\caption{Polarization-division STAR-DCM. (a) Schematic of metasurface element. (b) Conceptual illustration and details of the fabricated prototype.
 (c) Photograph of experimental setup and fabricated prototype.
 (d), (e), Representative measured scattering patterns, with blue and red color-coding indicating the two sides of the STAR-DCM. [used from \cite{Bao:2021pr} under  
		\href{https://creativecommons.org/licenses/by/4.0/}{CC BY 4.0}].}
	\label{fig:HW4}
\end{figure*}

Finally, Figure~\ref{fig:HW4} illustrates a reconfigurable polarization-division configuration operating at $11$ GHz \cite{Bao:2021pr}. The metasurface element, shown in Fig. \ref{fig:HW4}(a), exhibits a sandwich-like structure, consisting of three metal layers, two substrate layers and three PIN diodes, allowing switching between reflection and transmission modes (with phase modulation) under $x$- and $y$-polarized illumination, respectively. The conceptual illustration and some details of the fabricated $10\times20$-element prototype are shown in Fig. \ref{fig:HW4}(b), while the experimental setup is shown in Fig. \ref{fig:HW4}(c). This configuration enables full-space beam control via polarization-division, as exemplified by the scattering patterns in Figs.~\ref{fig:HW4}(d) and \ref{fig:HW4}(e).

Additionally, an implementation of STAR-DCMs through frequency division was proposed in \cite{Liu:2020rs}, with the programmable element designed so as to respond to reflection and transmission modes in two different frequency bands.

Hardware implementations of STAR-DCMs typically utilize PIN diodes in the microwave band. For higher frequencies, alternative implementations using graphene or phase-change materials (e.g., VO$_2$) can be explored. Compared with  metasurfaces operating solely in reflection or transmission, STAR-DCM hardware is considerably more complex, demanding  a synergistic approach involving discrete-signal processing and microelectronics. Large-scale STAR-DCMs typically integrate a large number of tunable devices, requiring numerous shift registers or a parallel FPGA control scheme in practice.

\section{ISAC with a STAR-DCM}
\label{sec:ISAC_STAR-DCM}

\begin{figure*}
	\centering
	\includegraphics[width=\linewidth]{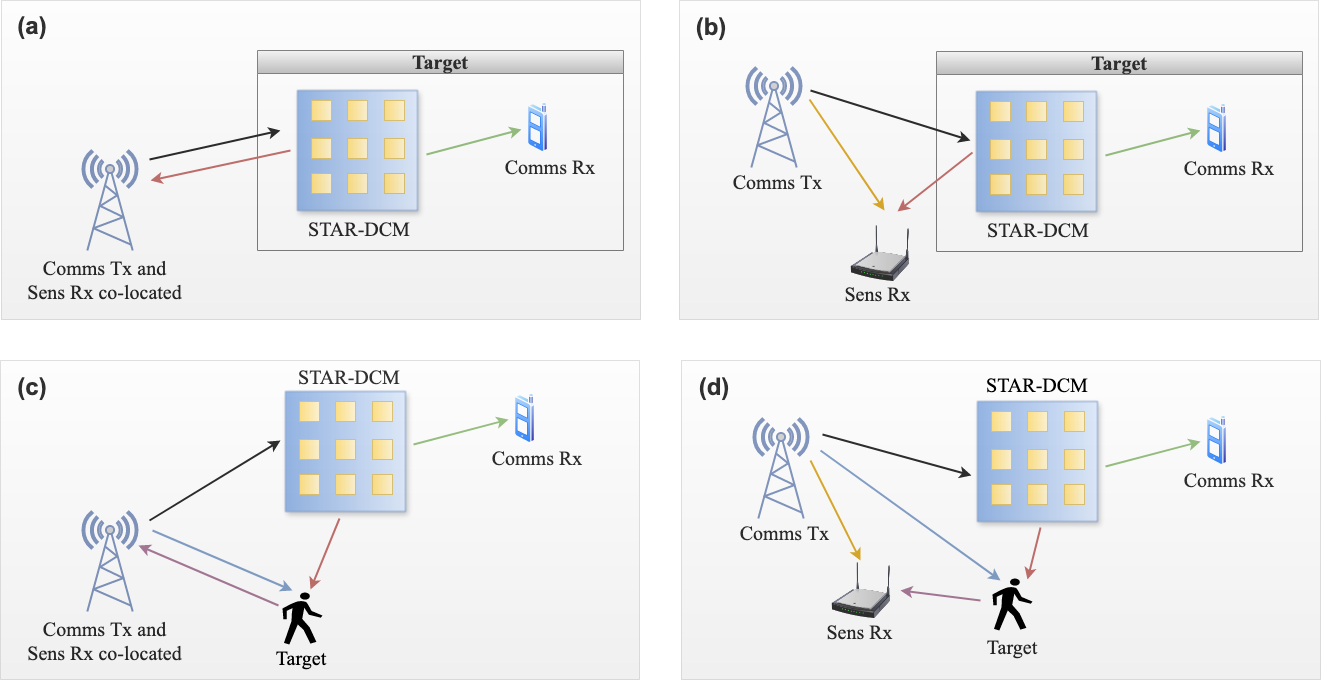}
	\caption{ISAC configurations with a STAR-DCM. (a) the STAR-DCM and the Comms Rx 
     are mounted on a target, and the Comms Tx and the Sens Rx are co-located (monostatic sensing of the STAR-DCM-plus-Comms-Rx target). (b) the STAR-DCM and the Comms Rx are mounted on a target, whereas the Comms Tx and the Sens Rx are placed at separate locations (bistatic sensing of the STAR-DCM-plus-Comms-Rx target). (c) the STAR-DCM and the Comms Rx are physically separated, whereas the Comms Tx and the Sens Rx are co-located (monostatic sensing of a third-party target). (d) the STAR-DCM and the Comms Rx are physically separated, and the Comms Tx and the Sens Rx are placed at separate locations (bistatic sensing of a third-party target).
     }
	\label{fig:ISAC_scenarios}
\end{figure*}

Various architectures may arise for implementing STAR-DCMs in upcoming ISAC systems, depending on the level of integration between sensing and communication functionalities. The schematic in Fig.~\ref{fig:ISAC_scenarios} illustrates four distinct configurations. This tutorial focuses specifically on point-to-point communications, involving a single communications transmitter (Comms Tx) and a single communications receiver (Comms Rx). We refer the reader to \cite{Wang.2023,Liu.2023,Eghbali.2024,Meng.2024} for the case of multiple communications users.
The network is simplified by assuming non-interfering point-to-point links, achieved through orthogonalization of user transmissions across temporal, spectral, spatial, angular, or code domains. The integration of STAR-DCMs into ISAC networks with NOMA is discussed in \cite{Sun.2024,Wang.2024,Xue_2024,Wei.2024}.

Figs.~\ref{fig:ISAC_scenarios}(a) and \ref{fig:ISAC_scenarios}(b) represent the category of {\em sensing-assisted communications} \cite{Meng.2024,Li.2024}. 
In this scenario, the target is represented by a crewed or unmanned vehicle. The metal panel (or a part thereof) on the top surface of the vehicle
is replaced with a STAR-DCM. A fraction of the signals impinging on the STAR-DCM
is partially transmitted through the vehicle to deliver data to a Comms Rx (e.g., an
onboard communication device or a mobile phone carried by a passenger), while the 
remaining fraction is reflected towards the sensor receiver (Sens Rx) to sense the vehicle. 
The Sens Rx collects data, enhancing network features in terms of: (i) Improved accuracy of channel estimation, reducing pilot transmission and associated overheads; (ii) Streamlined training overheads and minimized latency in beam alignment; (iii) Predictive capabilities for adaptive beam reconfiguration; (iv) Support for efficient resource allocation, including power optimization, bandwidth management, and cell handover facilitation by monitoring device velocity, azimuth angle, and heading direction; (v) Enhanced physical-layer security.
It is noteworthy that the use cases depicted in Figs.~\ref{fig:ISAC_scenarios}(a) and \ref{fig:ISAC_scenarios}(b) 
call for full-space control, which cannot be achieved by a reflecting-only DCM. Indeed, reflecting-only DCMs require sensing target and Comms Rx to be located 
on the same side of the DCM as the Comms Tx and the Sens Rx, hence achieving only half-space coverage.

Figs.~\ref{fig:ISAC_scenarios}(c) and \ref{fig:ISAC_scenarios}(d) illustrate {\em communication-assisted sensing} \cite{Wang.2023,Liu.2023}. This emerging concept enhances the sensing capabilities of a third-party target by utilizing signals reflected by the STAR-DCM. Examples include indoor localization of human subjects, like museum visitors or hospital patients, using communication signals between fixed control units and robotic entities. Besides providing a communication link within its transmission space, the STAR-DCM enhances object sensing accuracy in its reflection space. This paradigm is advantageous in scenarios with few transmitting terminals and non-line-of-sight environments. In future systems, communication-assisted sensing could offer unique opportunities, enabling sensing as a native service, often termed {\em sensing as a service}.
It is worth noting that, in Figs.~\ref{fig:ISAC_scenarios}(c) and \ref{fig:ISAC_scenarios}(d), 
the target is located on the reflection side of the DCM, while the Comms Rx
is on its transmission side. In this case, the DCM must operate in STAR mode. 
Conversely, if the target and the Comms Rx are on the same side, 
and hence only half-space coverage is required,
the DCM
operates in conventional transmitting/reflecting-only mode. However, this geographical restriction may not always be feasible in many ISAC 
scenarios, potentially limiting the flexibility and effectiveness of DCM for ISAC applications.

Focusing on the monostatic sensing configuration in Fig.~\ref{fig:ISAC_scenarios}(a), a segment of the incident signal on the STAR-DCM directs towards the Comms Rx for information conversion, while the remaining portion reflects back to the Sens Rx for beam prediction and tracking. In this scenario, the Comms Tx and Sens Rx are co-located within a multi-antenna ISAC transceiver, having $N$ transmit antennas and $M$ receive antennas. For simplicity, we consider the special case of a single-antenna Comms Rx.

In vehicle-to-everything (V2X) applications \cite{Meng.2024,Li.2024}, a common assumption is that both downlink (ISAC transceiver to vehicle) and uplink (vehicle to ISAC transceiver) share the same dominant line-of-sight (LoS) path.
Assuming LoS propagation, the time-varying state of the target includes direction $\boldsymbol{\iota}(t) \eqdef [\theta(t)], \phi(t)]^\trasp \in \mathbb{R}^2$, distance $d(t)$, and radial speed $v(t)$ in relation to the ISAC transceiver.
In practical scenarios, state parameters change slowly compared to the observation time interval \cite{Weijie.2021}. Thus, the observation interval is commonly divided into time slots of duration $\Delta T$ (see Fig.~\ref{fig:protocol}). Assuming that $\boldsymbol{\iota}(t) = \boldsymbol{\iota}_k \eqdef [\theta_k, \phi_k]^\trasp$, $d(t)=d_k$, and $v(t)=v_k$ remain constant over the $k$-th slot, where $k \in \mathbb{N}_0$, the {\em state} vector $\s_k \eqdef [\boldsymbol{\iota}_k^\trasp,d_k,v_k]^\trasp \in \mathbb{R}^4$ collects the main parameters of the target during the $k$-th time slot.

\subsection{STAR-DCM model}
\label{sec:meta-model}

The STAR-DCM comprises $L=L_\text{h} \times L_\text{v}$ reconfigurable elements arranged in a rectangular grid, indexed row-by-row by $\ell \in \{1,2, \ldots, L\}$. Assuming illumination by a linearly polarized, monochromatic plane wave with frequency $f_0>0$ and angular stability (i.e., weak dependence on the incidence direction \cite{Shab.2022}), the EM response of the $\ell$-th element is described by the {\em transmission coefficient} $\gamma_\text{T}^{(\ell, f_\text{T}, p_\text{T})}(t) \in \mathbb{C}$ controlling the fraction of the incident wave transmitted at the frequency $f_\text{T}$ along the polarization direction $p_\text{T} \in \{\text{h}, \text{v}\}$ (horizontal or vertical). Similarly, the {\em reflection coefficient} $\gamma_\text{R}^{(\ell, f_\text{R}, p_\text{R})}(t) \in \mathbb{C}$ dictates the fraction of the incident wave reflected at the frequency $f_\text{R}$ along the polarization direction $p_\text{R} \in \{\text{h},\text{v}\}$. Alternatively, right- and left-handed circularly polarized waves could be utilized \cite{Sun:2023tr}.
Discrete-time notations $\gamma_{\text{T},k}^{(\ell, f_\text{T}, p_\text{T})}$ and $\gamma_{\text{R},k}^{(\ell, f_\text{R}, p_\text{R})}$ are used when the coefficients remain constant over the $k$-th slot.
Neglecting cross-polarization effects and losses, power conservation imposes the constraints:
\barr
\big|\gamma_\text{T}^{(\ell, f_\text{T}, \text{h})}
(t)\big|^2 + 
\left|\gamma_\text{R}^{(\ell, f_\text{R}, \text{h})}
(t)\right|^2 & =1,
\label{eq:xx}
\\
\big|\gamma_\text{T}^{(\ell, f_\text{T}, \text{v})}
(t)\big|^2 + 
\big|\gamma_\text{R}^{(\ell, f_\text{R}, \text{v})}
(t)\big|^2 & =1 \:.
\label{eq:yy}
\earr
To account for the presence of losses, a factor less than one can be introduced on the right-hand side of \eqref{eq:xx} and \eqref{eq:yy}.
In line with Section~\ref{sec:EMdesign_implementation}, there 
are five  practical  options to be considered.

\subsubsection{Energy splitting}
The transmission and reflection coefficients remain constant 
over the $k$-th slot at the frequency $f_\text{T}=f_\text{R}=f_0$ 
and along the same polarization direction $p_\text{T}=p_\text{R}=p$. 
In this case, $\big|\gamma_{\text{T},k}^{(\ell, f_0, p)}\big| \neq 0$ and $|\gamma_{\text{R},k}^{(\ell, f_0, p)}| \neq 0$
are coupled through either \eqref{eq:xx} or \eqref{eq:yy}.

\subsubsection{Mode switching}
With $f_\text{T}=f_\text{R}=f_0$ 
and $p_\text{T}=p_\text{R}=p$, during the $k$-th slot,   
the $\ell$-th element operates either in transmission 
mode, i.e., $\big|\gamma_{\text{T},k}^{(\ell, f_0, p)}\big|=1$ 
and $\gamma_{\text{R},k}^{(\ell, f_0, p)}=0$,  or in reflection mode, i.e., $\big|\gamma_{\text{R},k}^{(\ell, f_0, p)}\big|=1$ and $\gamma_{\text{T},k}^{(\ell, f_0, p)}=0$.   
The magnitudes of the transmission and reflection coefficients
are straightforwardly decoupled.

\subsubsection{Time division}
With $f_\text{T}=f_\text{R}=f_0$ 
and $p_\text{T}=p_\text{R}=p$, 
the $\ell$-th element switches between
transmission and  reflection modes 
in orthogonal time intervals within 
the $k$-th slot. Specifically, at the time instant $t=t_1$, 
$\big|\gamma_\text{T}^{(\ell, f_0, p)}
(t_1)\big| =1$ and 
$\gamma_\text{R}^{(\ell, f_0, p)}
(t_1) =0$, whereas, when $t=t_2 \neq t_1$, 
$\big|\gamma_\text{R}^{(\ell, f_0, p)}
(t_2)\big| =1$, and 
$\gamma_\text{T}^{(\ell, f_0, p)}
(t_2) =0$.
The magnitudes of the transmission and reflection coefficients
are decoupled in time.

\subsubsection{Polarization division}
With $f_\text{T}=f_\text{R}=f_0$, during the 
$k$-th slot,  
the $\ell$-th element only operates in transmission [reflection]
mode for the h-polarized incident wave, i.e., 
$\gamma_{\text{R},k}^{(\ell, f_0, \text{h})}=0$ [$\gamma_{\text{T},k}^{(\ell, f_0, \text{h})} =0$], 
and  in reflection [transmission]
mode for the v-polarized incident wave, i.e., 
$\gamma_{\text{T},k}^{(\ell, f_0, \text{v})} =0$ [$\gamma_{\text{R},k}^{(\ell, f_0, \text{v})} =0$].
The magnitudes of the transmission and reflection coefficients
are decoupled in polarization.

\subsubsection{Frequency division}
With $p_\text{T}=p_\text{R}=p$, during the 
$k$-th slot,  the $\ell$-th element switches between
transmission and  reflection modes 
in orthogonal frequency intervals, i.e., 
at the illumination frequency $f_0=f_1$, 
$\big|\gamma_{\text{T},k}^{(\ell, f_1, p)}\big| =1$ and 
$\gamma_{\text{R},k}^{(\ell, f_1, p)}=0$, whereas, 
when $f_0=f_2 \neq f_1$, $\big|\gamma_{\text{R},k}^{(\ell, f_2, p)}\big| =1$
and 
$\gamma_{\text{T},k}^{(\ell, f_2, p)} =0$.
The magnitudes of the transmission and reflection coefficients
are decoupled in frequency.

\subsection{Signal models}
\label{sec:sig-model}

The received sensing and communication signals
are explained using the simplified 
cascaded channel model \cite{Xu.2022,Mu.2022}.
Here, we assume the STAR-DCM 
located in the far-field region of the Comms Tx,
and both the Comms Rx and the Sens Rx 
in the far-field region
of the STAR-DCM.
Moreover, we make the common assumption that the
one-sided bandwidth $B$ of the ISAC signal is much smaller than its carrier
frequency $f_0$, ensuring that the response of the arrays and
the DCM is essentially
constant within the frequency interval $(f_0-B/2, f_0+B/2)$.

\subsubsection{Received sensing signal}

At the $k$-th time slot,
the baseband signal $\y_{\text{R},k}(t)$
received by the Sens Rx can be expressed as 
\be
\y_{\text{R},k}(t)= \G_k(t; \s_k) \, \Gammab_{\text{R},k}(t) \,  
\F_k(t; \s_k) \, \x_k(t-\tau_{\text{R},k}) + \w_{\text{R},k}(t),
\label{eq:ys}
\ee
where $\F_k(t; \s_k) \in \mathbb{C}^{L \times M}$
and $\G_k(t; \s_k) \in \mathbb{C}^{M \times L}$ 
model the Comms-Tx-to-STAR-DCM and 
STAR-DCM-to-Sens-Rx channels (including
large-scale fading effects, as well as responses of the arrays and DCM), respectively,
$\Gammab_{\text{R},k}(t) \eqdef \diag[\gammab_{\text{R},k}(t)]$, with  
$\gammab_{\text{R},k}(t) \in \mathbb{C}^{ L}$ containing 
the reflection coefficient of the STAR-DCM,  
$\x_k(t) \in \mathbb{C}^N$ is the transmit vector
acting as both sensing waveform and 
information-bearing signal,
$\tau_{\text{R},k}$ is the round-trip delay, and 
$\w_{\text{R},k}(t)$ is a zero-mean white Gaussian vector accounting
for thermal noise with 
covariance matrix $\sigma_{w_\text{R}}^2 \, \mathbf{I}_M$.
Notably, both 
$\F_k(t; \s_k)$ and $\G_k(t; \s_k)$ are time-varying, since they also encompass
the carrier frequency shift induced by the Doppler effect. Moreover, 
$\F_k(t; \s_k)$ and $\G_k(t; \s_k)$ explicitly depend on the state vector $\s_k$. 
The signal received at the antenna array of the Sens Rx is spatially combined 
for sensing. The signal at the output of the combiner is given by 
$z_{\text{R},k}(t)= \vb_k^\herm(t) \, \y_{\text{R},k}(t)$, 
where $\vb_k(t) \in \mathbb{C}^M$ is the beamforming vector
at the Sens Rx.

\subsubsection{Received communication signal}

The transmission channel between the STAR-DCM and the Comms Rx
can be reasonably assumed to be time-invariant over the observation 
interval and frequency-flat (i.e., the delay channel 
spread is small relative to the inverse signal bandwidth), since the 
STAR-DCM and the Comms Rx
remain relatively stationary within the mobile target,
and the distance between them is short.
Moreover, links between the Comms Tx and Comms Rx can be assumed 
to be negligible in the case of severe penetration losses. 
Henceforth, at the $k$-th time slot,
the baseband signal $y_{\text{T},k}(t)$
received by the Comms Rx is only given by the 
wave transmitted by the STAR-DCM, which can be 
written as 
\be
y_{\text{T},k}(t)= \cb_k^\trasp \, \Gammab_{\text{T},k}(t) \,  
\F_k(t; \s_k) \, \x_k(t-\tau_{\text{T},k}) + w_{\text{T},k}(t),
\label{eq:yr}
\ee
where  $\cb_k \eqdef [c_{k,1},c_{k,2}, \ldots, c_{k,L}]^\trasp \in \mathbb{C}^L$ represents the STAR-DCM-to-Comms-Rx
channel (including fading effects and the DCM response), 
$\Gammab_{\text{T},k}(t) \eqdef \diag[\gammab_{\text{T},k}(t)]$, with  
$\gammab_{\text{T},k}(t) \in \mathbb{C}^{ L}$ containing 
the transmission coefficient of the STAR-DCM,  
$\tau_{\text{T},k}$ is the overall delay,
and $w_{\text{T},k}(t)$ is a zero-mean white Gaussian thermal noise 
with variance $\sigma_{w_\text{T}}^2$.

\begin{figure*}
	\centering
	\includegraphics[width=0.8\linewidth]{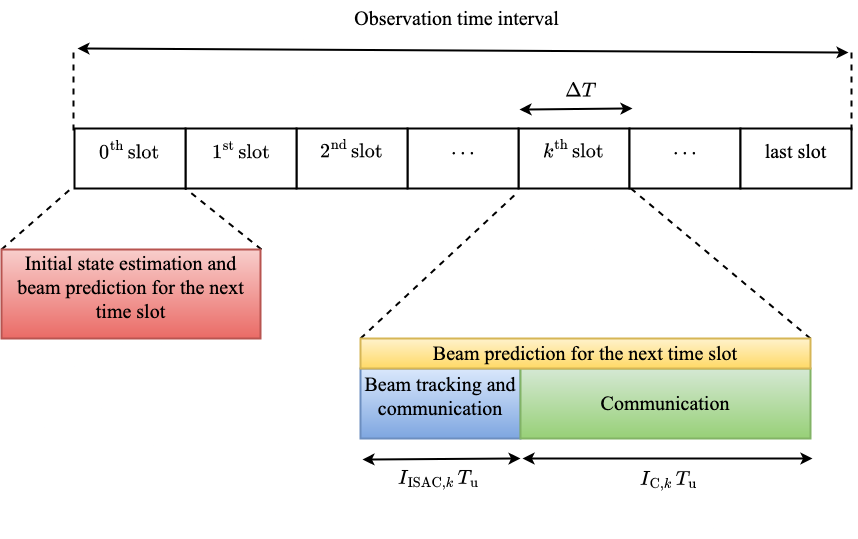}
	\caption{Illustration of a protocol with sensing-assisted beam 
              prediction and tracking. $T_\text{u}$ denotes the symbol period of the information-bearing digitally-modulated signal transmitted by the Comms Tx,
              while $I_{\text{ISAC},k} \in \mathbb{N}$ and $I_{\text{C},k} \in \mathbb{N}$ are the number of symbol intervals devoted to ISAC and communication-only processes, respectively.}
	\label{fig:protocol}
\end{figure*}

\subsection{Beam prediction and tracking based on STAR-DCM}
\label{sec:beam}

Beam prediction and tracking are crucial operations 
to ensure beamforming gains 
in fast-changing channels for the Comms Tx. 
Let us consider the transmit beamforming strategy 
\be
\x_k(t) = \bb_k(t) \,  u_k(t),
\ee
where $\bb_k(t) \in \mathbb{C}^M$ is the beamformer used for steering the beam 
to the desired direction during the $k$-th time slot, 
and $u_k(t)$ is the corresponding  unit-power data stream 
intended for the Comms Rx with a symbol period $T_\text{u}$. To steer the transmitted 
signal $u_k(t)$ towards the direction of the mobile target, knowledge
of the state vector $\s_{k}$  is required at the Comms Tx.
This information is provided by the Sens Rx. 
Fig.~\ref{fig:protocol} illustrates 
a possible beam prediction and tracking protocol involving 
a STAR-DCM \cite{Meng.2024}. The protocol 
is initialized by letting the Sens Rx estimate the parameters 
of the target that enters into the coverage of interest.

In the initial state-estimation step ($0$-th slot), the Sens Rx derives
an initial estimate $\widehat{\s}_0$ of the state vector $\s_0$
from the reflected signal $\y_{\text{R},0}(t)$ by configuring the DCM to be in a totally reflective state (i.e., 
$\gammab_{\text{T},0}(t)=0$). The state estimate
can be obtained using standard
training-based estimation methods. In this approach, the Comms Tx transmits a pilot signal
known to the Sens Rx, and simultaneously, 
the reflection response 
$\gammab_{\text{R},0}(t)$ of the DCM is 
set to be in a training configuration 
known to the Sens Rx.

\begin{figure*}
	\centering
	\includegraphics[width=0.8\linewidth]{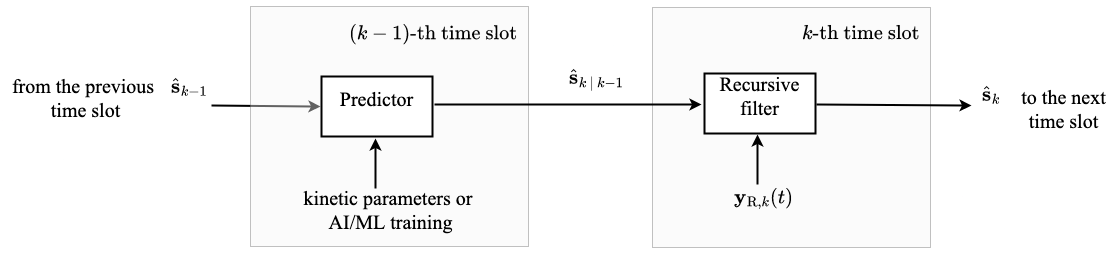}
	\caption{Prediction and tracking of the target state vector.}
	\label{fig:prediction_tracking}
\end{figure*}

As illustrated in Fig.~\ref{fig:protocol}, beam prediction 
is performed at the Sens Rx on a slot-by-slot basis. The goal of the prediction process in
the $(k-1)$-th time slot is to predict the state of 
the target at the $k$-th slot
using the estimate $\widehat{\s}_{k-1}$ of $\s_{k-1}$ obtained in 
the previous $(k-2)$-th slot. 
This procedure is schematically illustrated in Fig.~\ref{fig:prediction_tracking}.
The predictor can be designed through either model-based or 
data-driven approaches. Model-based prediction relies on
the kinetic model of the mobile target \cite{Weijie.2021}, whereas  
data-driven prediction utilizes  
artificial intelligence (AI) or machine learning (ML)
techniques \cite{Li.2023}. In what follows, we denote  
$\widehat{\s}_{k \,|\, k-1} \eqdef  
[\widehat{\boldsymbol{\iota}}_{k \,|\, k-1}^\trasp, \widehat{d}_{k \,|\, k-1}, 
\widehat{v}_{k \,|\, k-1}]^\trasp$ as the output of the predictor during the $(k-1)$-th slot.

The prediction $\widehat{\s}_{k \,|\, k-1}$ serves as a {\em coarse estimate} 
of the target state vector. The Comms Tx uses this estimate to initially steer the beam 
at the beginning of the $k$-th time slot, i.e., at the time epoch $t_k =k \, \Delta T$. 
A  simple transmit beamforming strategy involves aligning $\bb_k(t_k)$ to the 
coarsely predicted Comms-Tx-to-STAR-DCM LoS direction by setting
\be
\bb_k(t_k) = \bb_{k,1} \eqdef \frac{\ab_\text{Comms-Tx}(\widehat{\boldsymbol{\iota}}_{k \,|\, k-1})}
{\| \ab_\text{Comms-Tx}(\widehat{\boldsymbol{\iota}}_{k \,|\, k-1})\|_2},
\label{eq:b-1}
\ee
where $\ab_\text{Comms-Tx}(\cdot) \in \mathbb{C}^N$ is 
the  array response vector of the Comms Tx.
Similarly, the receive beamforming is designed by matching 
\be
\vb_k(t_k) \eqdef \vb_{k,1} \eqdef \frac{\ab_\text{Sens-Rx}(\widehat{\boldsymbol{\iota}}_{k \,|\, k-1})}
{\| \ab_\text{Sens-Rx}(\widehat{\boldsymbol{\iota}}_{k \,|\, k-1})\|_2}, 
\label{eq:v-1}
\ee
with the array response vector $\ab_\text{Sens-Rx}(\cdot) \in \mathbb{C}^M$ of the Sens Rx
along the direction $\widehat{\boldsymbol{\iota}}_{k \,|\, k-1}$.

The beam tracking procedure  aims at refining
the transmit beam direction during the $k$-th time slot 
so as to enhance communication
quality. As shown in Fig.~\ref{fig:prediction_tracking}, 
this is achieved by implementing a {\em recursive} estimation procedure
at the Sens Rx. The goal is 
to provide a {\em fine estimate} $\widehat{\s}_{k} \eqdef 
[\widehat{\boldsymbol{\iota}}_{k}^\trasp, \widehat{d}_{k}, 
\widehat{v}_{k}]^\trasp$
of the target state vector based on the  
coarse estimate $\widehat{\s}_{k \,|\, k-1}$ 
and the reflected signal $\y_{\text{R},k}(t)$ from the STAR-DCM.
For this purpose, similarly to conventional (i.e., without DCM)
ISAC systems, an extended Kalman filtering can be used 
in conjunction with standard high-resolution direction-finding algorithms 
and delay-Doppler matched filtering techniques \cite{Weijie.2021}.
The protocol consists of dividing the time slot $k$ 
into two sub-slots \cite{Meng.2024} (see Fig.~\ref{fig:protocol}). 
In the first one, the number 
of symbol intervals allocated to joint beam tracking and communication 
is $I_{\text{ISAC},k}$, whereas the 
remaining $I_{\text{C},k}$ symbol intervals are devoted to
information transfer only.  
In the first sub-slot,
the Comms Tx and the Sens Rx keep the beamformer 
fixed to the initial value \eqref{eq:b-1} and 
\eqref{eq:v-1}, respectively,  
and the DCM operates in STAR mode, 
according to one of
the mechanisms in Subsection~\ref{sec:meta-model},
with  
reflection and transmission responses 
$\gammab_{\text{R},k}(t)=\gammab_{\text{R},k,1} \neq \mathbf{0}_L$ and 
$\gammab_{\text{T},k}(t)=\gammab_{\text{T},k,1}\neq \mathbf{0}_L$. 
For $t \in [k \, \Delta T, k \, \Delta T+I_\text{ISAC}\, T_\text{u})$,
the fine estimate $\widehat{\s}_{k}$ is obtained by jointly processing 
$\widehat{\s}_{k \,|\, k-1}$ and \eqref{eq:ys}, 
while the Comms Rx is receiving information according to
\eqref{eq:yr}. 

In the second sub-slot, the Comms Tx refines  
its matched-filter beamformer, by setting 
\be
\bb_k(t)  = \bb_{k,2} \eqdef \frac{\ab_\text{Comms-Tx}(\widehat{\boldsymbol{\iota}}_{k})}
{\| \ab_\text{Comms-Tx}(\widehat{\boldsymbol{\iota}}_{k})\|_2}
\label{eq:b-2}
\ee
for $t \in [k \, \Delta T+I_\text{ISAC}\, 
T_\text{u}, (k+1) \, \Delta T)$ and, at the same time, 
the DCM is switched to a totally transmission configuration, i.e., 
$\gammab_{\text{R},k}(t)=\gammab_{\text{R},k,2}=\mathbf{0}_L$ and 
$\gammab_{\text{T},k}(t)=\gammab_{\text{T},k,2} \neq \mathbf{0}_L$. 

The outlined protocol does not require uplink feedbacks, as the signal reflected by the STAR CDM inherently provides this feedback. Dedicated downlink pilots are necessary only during 
the initial estimation step. 
An implicit assumption  here is that the 
STAR-DCM-plus-Comms-Rx target 
is frame-synchronous with the Comms Tx/Sens Rx. 
This assumption is common 
in all works dealing with DCM-enabled ISAC networks, so it is not particularly
restrictive to STAR-DCMs. 
Frame synchronization can be established during 
the initial estimation step and maintained
throughout the protocol.
However, such a task remains an open challenge in signal processing.

\section{Optimization of ISAC with a STAR-DCM}
\label{sec:optimization}

The effectiveness of an ISAC system relies on achieving both 
sensing accuracy and reliable communication. 
The {\em posterior Cram\'{e}r–Rao bound (PCRB)} \cite{Van-Trees} establishes 
a lower bound on the variance of unbiased estimators for 
each entry of the target state vector. Simultaneously,  communication 
performance is assessed by the achievable {\em information data rate}.

The PCRB on the estimation error can be expressed as
\be
\mathbf{P} \eqdef \Es\left[ \left(\widehat{\s}_{k}-\s_{k}\right) \, \left(\widehat{\s}_{k}-\s_{k}\right)^\herm \right] \succeq \mathbf{J}^{-1},
\label{eq:PCRB}
\ee
where the inequality means that the difference
$\mathbf{P}-\mathbf{J}^{-1}$ is a positive semidefinite matrix, 
whereas $\mathbf{J}$ is the Fisher information matrix defined as 
\be
\mathbf{J} \eqdef  \left[ -\frac{\partial^2  \log p(\y_{\text{R},k}(t_k),\s_{k})}{\partial \s_{k}^2}\right],
\label{eq:FIM}
\ee
with $p(\y_{\text{R},k}(t_k),\s_{k})$ denoting the joint probability density of the pair 
$\{\y_{\text{R},k}(t_k),\s_{k}\}$,
provided that the expectations and derivatives in \eqref{eq:PCRB} and \eqref{eq:FIM} exist.
The calculation of the PCBR for the problem at hand is mainly complicated by 
the fact that  $\s_{k}$ is a nonlinear function of $\s_{k-1}$,
 making it difficult to analytically derive the distribution of $\s_{k}$.
To avoid overwhelming the reader with cumbersome mathematical derivations, we note
that the PCRB for the estimates  
$\widehat{\boldsymbol{\iota}}_{k}$, $\widehat{d}_{k}$, 
and $\widehat{v}_{k}$, which collectively determine the estimate
$\widehat{\s}_{k}$ of the target state vector during the 
$k$-th time slot, is inversely proportional to both 
the number of measurements $I_{\text{ISAC},k}$ of the
reflected signal by the STAR-DCM and 
the  signal-to-noise ratio (SNR) of the Sens Rx beamformer's output 
$z_{\text{R},k}(t)= \vb_k^\herm(t) \, \y_{\text{R},k}(t)$. The constants of proportionality 
depend on the system configuration, signal designs, 
and specific estimation algorithms employed.
This observation allows us to streamline the presentation of the system
optimization while providing insights into
the benefits of STAR-DCM-aided ISAC systems. 
For a detailed derivation of the PCRB, we refer the reader to \cite{Meng.2024}.

Utilizing the predicted $\widehat{\boldsymbol{\iota}}_{k \,|\, k-1}$
of the STAR-DCM-plus-Comms-Rx target
direction obtained in the $(k-1)$-th slot
(see Fig.~\ref{fig:prediction_tracking}), 
according to \eqref{eq:ys}, 
the SNR of the signal $z_{\text{R},k}(t)$ 
during the first sub-slot is given by 
\be  
\text{SNR}_{k,1}= 	
\frac{\pot_{\text{R},k} \, \big|\gain_{k,1}^{\text{Sens-Rx}}\big|^2 \, 
\big |\ab_\text{DCM}^\herm(-\bu_k) \, 
\Gammab_{\text{R},k,1}\, \ab_\text{DCM}(\bu_k)\big|^2 \, 
\big|\gain_{k,1}^\text{Comms-Tx}\big|^2}{\sigma_{w_\text{R}}^2}, 
\label{eq:snr}
\ee  
where  
$\pot_{\text{R},k}$ is the received power at the Sens Rx (encompassing
transmit power and round-trip path loss), 
$\big|\gain_{k,1}^\text{Sens-Rx}\big|^2= \big|\vb_{k,1}^\herm \, 
\ab_\text{Sens-Rx}(\boldsymbol{\iota}_k)\big|^2 \le M$
represents the array gain of the Sens Rx, 
with $\vb_{k,1}$ given by \eqref{eq:v-1}, 
$\ab_\text{DCM}(\cdot) \in \mathbb{C}^L$ models the DCM response,  
$\bu_k \in \mathbb{R}^3$ collects the 
cosines of the direction $\boldsymbol{\iota}_k$, 
$\Gammab_{\text{R},k,1}=\diag[\gammab_{\text{R},k,1}]$, 
and, finally, the array gain of the Comms Tx in the first sub-slot is denoted by
$\big|\gain_{k,1}^\text{Comms-Tx}\big|^2= 
\big|\ab_\text{Comms-Tx}^\herm(\boldsymbol{\iota}_k) \, \bb_{k,1}\big|^2 \le N$,
with $\bb_{k,1}$ given by \eqref{eq:b-1}.
The losses in transmit and receive 
array gain, denoted as $N-\big|\gain_{k,1}^\text{Comms-Tx}\big|^2$
and $M-\big|\gain_{k,1}^\text{Sens-Rx}\big|^2$, respectively,
depend on the accuracy of the prediction 
$\widehat{\boldsymbol{\iota}}_{k \,|\, k-1}$. 
As $\big \|\widehat{\boldsymbol{\iota}}_{k \,|\, k-1}-\boldsymbol{\iota}_k \big\|_2
\to 0$,
the array gains $\big|\gain^\text{Comms-Tx}_{k,1}\big|^2$ and 
$\big|\gain^\text{Sens-Rx}_{k,1}\big|^2$ approach their maximum value $N$ and $M$, respectively.

Considering \eqref{eq:yr}, the achievable rate in the initial sub-slot of the $k$-th slot is given by
\be
\rate_{k,1} = \log_2\left(1+ \frac{\pot_{\text{T},k}\, 
\big| \cb_k^\trasp \, \Gammab_{\text{T},k,1}\, \ab_\text{DCM}(\boldsymbol{\iota}_k)
\big|^2 \, \big|\gain_{k,1}^\text{Comms-Tx}\big|^2}{\sigma_{w_\text{T}}^2}\right)
\quad \text{(in bits/s/Hz)},
\label{eq:rate-1}
\ee
where $\pot_{\text{T},k}$ is  
the received power at the Comms Rx (encompassing
transmit power and overall path loss), and 
$\Gammab_{\text{T},k,1}=\diag[\gammab_{\text{T},k,1}]$.
On the other hand, the achievable rate in the second 
sub-slot of the $k$-th slot is expressed as
\be
\rate_{k,2} = \log_2\left(1+ \frac{\pot_{\text{T},k}\, 
\big| \cb_k^\trasp \, \Gammab_{\text{T},k,2} \, \ab_\text{DCM}(\boldsymbol{\iota}_k)
\big|^2 \, \big|\gain_{k,2}^\text{Comms-Tx}\big|^2}{\sigma_{w_\text{T}}^2}\right)
\quad \text{(in bits/s/Hz)},
\label{eq:rate-2}
\ee
where $\big|\gain_{k,2}^\text{Comms-Tx}\big|^2= 
\big|\ab_\text{Comms-Tx}^\herm(\boldsymbol{\iota}_k) \, \bb_{k,2}\big|^2 \le N$
is  the array gain of the Comms Tx in the second sub-slot, 
with $\bb_{k,2}$ given by \eqref{eq:b-2},
and $\Gammab_{\text{T},k,2}=\diag[\gammab_{\text{T},k,2}]$,
which approaches its maximum value 
$N$ as $\big\|\widehat{\boldsymbol{\iota}}_{k}-\boldsymbol{\iota}_k \big\|_2 \to 0$,

\subsection{Optimization of the STAR-DCM}
\label{sec:opt-prob}

To leverage the dual-function capability of ISAC, various 
optimization problems can be formulated based
on the design objective and                             
on the type of STAR mechanism  
(see Subsection~\ref{sec:meta-model}). 
Broadly, the optimization scope might involve maximizing 
sensing performance while meeting a specified 
quality-of-service requirement for the Comms Rx. 
Conversely, one could aim to maximize the 
information rate for the Comms Rx while ensuring a certain
level of accuracy in estimating the 
unknown state parameters. The latter is particularly relevant for sensing-assisted communications. 
The selection of the STAR mechanism
significantly influences the mutual coupling between the transmission and
reflection coefficients.  

Let $\gamma_{\text{R},k,1}^{(\ell)}$, $\gamma_{\text{T},k,1}^{(\ell)}$, 
and $\gamma_{\text{T},k,2}^{(\ell)}$ denote the $\ell$-th entry of $\gammab_{\text{R},k,1}$,
$\gammab_{\text{T},k,1}$, and
$\gammab_{\text{T},k,2}$, respectively, for $\ell \in \{1,2,\ldots, L\}$.
In scenarios involving mode switching, as well as 
time, frequency and polarization division, the magnitudes of 
$\gamma_{\text{R},k,1}^{(\ell)}$ and $\gamma_{\text{T},k,1}^{(\ell)}$
are decoupled, simplifying the optimization process. 
STAR-DCMs with mode switching
exhibit limited beamforming gains as only a subset of the elements
is selected for transmission or reflection. On the other hand,
ISAC systems employing STAR-DCMs with time or frequency division 
introduce stringent synchronization requirements, leading to  higher 
hardware implementation complexity.
Compared to time and frequency division, polarization division separates 
the design of $\gamma_{\text{R},k,1}^{(\ell)}$ and $\gamma_{\text{T},k,1}^{(\ell)}$
by alleviating the burden of DCM synchronization. Nevertheless, the full potential
of STAR-DCMs with polarization division in ISAC networks remains to be fully understood. 

To explore the inherent trade-off between sensing and communication 
performance in the monostatic sensing configuration illustrated in 
Fig.~\ref{fig:ISAC_scenarios}(a), we focus on a STAR-DCM employing 
energy splitting. Although this configuration couples the design of 
$\gamma_{\text{R},k,1}^{(\ell)}$ and $\gamma_{\text{T},k,1}^{(\ell)}$, 
it offers considerable flexibility for communication system design.
We preliminary discuss the maximization
of $\text{SNR}_{k,1}$ defined in \eqref{eq:snr}
with respect to $\Gammab_{\text{R},k,1}$, which directly enhances sensing capability by resulting in a lower PCRB. 
To this aim, it is convenient to observe 
that 
$\big |\ab_\text{DCM}^\herm(-\bu_k) \, 
\Gammab_{\text{R},k,1}\, \ab_\text{DCM}(\bu_k)\big|^2= \big| \boldsymbol{\alpha}_\text{DCM}^\herm(\boldsymbol{\iota}_k) 
\, \gammab_{\text{R},k,1}\big|^2$, where 
\be
\boldsymbol{\alpha}_\text{DCM}(\boldsymbol{\iota}_k) \eqdef [e^{-j \, 2\, 
\bk^\trasp(\boldsymbol{\iota}_k) \, \pb_1}, 
e^{-j \, 2 \, \bk^\trasp(\boldsymbol{\iota}_k) \, \pb_2}, \ldots,
e^{-j \, 2 \, \bk^\trasp(\boldsymbol{\iota}_k) \, \pb_L}]^\trasp \in \mathbb{C}^L,
\label{eq:alpha}
\ee
with $\bk(\boldsymbol{\iota}_k)=\frac{2 \pi}{\lambda_0} \, \bu_k$ representing the 
wavevector ($\lambda_0$ is the wavelength corresponding to the carrier 
frequency $f_0$) and $\pb_\ell$ denoting the spatial location of the $\ell$-th 
element of the DCM. The Cauchy-Schwarz inequality implies that 
$\big| \boldsymbol{\alpha}_\text{DCM}^\herm(\boldsymbol{\iota}_k) 
\, \gammab_{\text{R},k,1}\big|^2$ is maximized
when $\gammab_{\text{R},k,1}= \sqrt{\lambda_k} \, \boldsymbol{\alpha}_\text{DCM}(\boldsymbol{\iota}_k)$ for some $0 < \lambda_k \le 1$,
and its maximum value is $L^2$. 
Hence, achieving the maximum SNR during the first sub-slot of the
$k$-th slot involves ensuring that: (i) the phase of the $\ell$-th reflection coefficient satisfies 
$\measuredangle \gamma_{\text{R},k,1}^{(\ell)}=- 2 \,
\bk^\trasp(\boldsymbol{\iota}_k) \, \pb_\ell$; 
(ii) all the reflection coefficients of the DCM have the same
magnitude, i.e., $\big |\gamma_{\text{R},k,1}^{(\ell)} \big|^2=\lambda_k$,
for each $\ell \in \{1,2,\ldots,L\}$. 
Since the prediction 
$\widehat{\boldsymbol{\iota}}_{k \,|\, k-1}$ is only available 
at the beginning of the $k$-th time slot, the vector 
$\gammab_{\text{R},k,1}$ is designed by exploiting this predicted 
result, thus obtaining 
\be
\gammab_{\text{R},k,1}= \sqrt{\lambda_k} \, \boldsymbol{\alpha}_\text{DCM}(\widehat{\boldsymbol{\iota}}_{k \,|\, k-1}), 
\label{eq:refl-vec}
\ee
whose corresponding SNR can be derived from \eqref{eq:snr} as follows: 
\be  
\text{SNR}_{k,1}^\text{max}= 	
\frac{\lambda_k \, \pot_{\text{R},k} \, \big|\gain_{k,1}^{\text{Sens-Rx}}\big|^2 \, 
\big|\gain_{k,1}^\text{DCM}\big|^2 \, 
\big|\gain_{k,1}^\text{Comms-Tx}\big|^2}{\sigma_{w_\text{R}}^2}, 
\label{eq:snr-max}
\ee  
where $\big|\gain_{k,1}^\text{DCM}\big|^2= \big |
\boldsymbol{\alpha}_\text{DCM}^\herm(\boldsymbol{\iota}_k) \, 
\boldsymbol{\alpha}_\text{DCM}(\widehat{\boldsymbol{\iota}}_{k \,|\, k-1}) \big|^2$
is referred to as the DCM reflection gain.
The design in \eqref{eq:refl-vec} results in a DCM reflection gain loss of 
$L^2-\big|\gain_{k,1}^\text{DCM}\big|^2$, which decreases with increasing prediction accuracy, i.e., as $\widehat{\boldsymbol{\iota}}_{k \,|\, k-1} \to \boldsymbol{\iota}_k$. 
The continuous phases in \eqref{eq:refl-vec}
are subject to a quantization process that transforms each 
$\measuredangle \gamma_{\text{R},k,1}^{(\ell)}=-2 \,
\bk^\trasp(\widehat{\boldsymbol{\iota}}_{k \,|\, k-1}) \, \pb_\ell$ 
into one of a finite set $\mathcal{F}_\text{R}$ of prescribed values
(based on the digital code used for encoding the reflected phases).
The DCM reflection gain loss due to the quantization error can be 
mitigated by employing a set $\mathcal{F}_\text{R}$ with a sufficiently large cardinality,
and it is neglected in the subsequent discussion.

Enforcing both the power
conservation constraint $\big |\gamma_{\text{T},k,1}^{(\ell)} \big|^2 +\big |\gamma_{\text{R},k,1}^{(\ell)} \big|^2=1$ and 
the condition  $\big |\gamma_{\text{R},k,1}^{(\ell)} \big|^2=\lambda_k$, it follows that, during the first sub-slot of the
$k$-th slot, the transmission coefficients of the DCM also share the same
magnitude, i.e., $\big |\gamma_{\text{T},k,1}^{(\ell)} \big|^2= 1-\lambda_k$, 
for any element index $\ell \in \{1,2,\ldots,L\}$.
Applying similar reasoning as in \eqref{eq:refl-vec}, the  
corresponding phases that maximize 
$\big| \cb_k^\trasp \, \Gammab_{\text{T},k,1}\, \ab_\text{DCM}(\boldsymbol{\iota}_k)
\big|^2$ in \eqref{eq:rate-1}, with $\boldsymbol{\iota}_k$ replaced by 
$\widehat{\boldsymbol{\iota}}_{k \,|\, k-1}$, are given by 
\be
\measuredangle \gamma_{\text{T},k,1}^{(\ell)}=- \,
\bk^\trasp(\widehat{\boldsymbol{\iota}}_{k \,|\, k-1}) \, \pb_\ell -
\measuredangle c_{k,\ell} \: .
\label{eq:tx-phase-1}
\ee
Similarly, during the second sub-slot of the
$k$-th slot, optimizing the rate
\eqref{eq:rate-2} amounts to maximizing 
$\big| \cb_k^\trasp \, \Gammab_{\text{T},k,2} \, \ab_\text{DCM}(\boldsymbol{\iota}_k)
\big|^2$, with $\boldsymbol{\iota}_k$ replaced by 
$\widehat{\boldsymbol{\iota}}_{k}$. To do this, 
the transmission coefficients of the DCM are refined 
by setting $\big |\gamma_{\text{T},k,2}^{(\ell)} \big|^2 =1$  and
\be
\measuredangle \gamma_{\text{T},k,2}^{(\ell)}=- \,
\bk^\trasp(\widehat{\boldsymbol{\iota}}_{k}) \, \pb_\ell -
\measuredangle c_{k,\ell}
\label{eq:tx-phase-2}
\ee
for each $\ell \in \{1,2,\ldots,L\}$. Both 
\eqref{eq:tx-phase-1} and \eqref{eq:tx-phase-2}
must be quantized to the feasible set $\mathcal{F}_\text{T}$
of values for the transmission-coefficient phases
(depending on the digital codes used for encoding the transmitted phases).
By substituting \eqref{eq:tx-phase-1}--\eqref{eq:tx-phase-2} in 
\eqref{eq:rate-1}--\eqref{eq:rate-2} and neglecting quantization effects
(a reasonable assumption for a sufficiently large cardinality of $\mathcal{F}_\text{T}$), we obtain
\barr
\rate_{k,1}^\text{max} & =
\log_2\left(1+ \frac{(1-\lambda_k) \, \pot_{\text{T},k}\,  
\big|\gain_{k,1}^\text{Comms-Rx}\big|^2 \, \big|\gain_{k,1}^\text{Comms-Tx}\big|^2}{\sigma_{w_\text{T}}^2}\right),
\label{eq:Rk1-max}
\\
\rate_{k,2}^\text{max} & = 
\log_2\left(1+ \frac{\pot_{\text{T},k}\,  
\big|\gain_{k,2}^\text{Comms-Rx}\big|^2 \, \big|\gain_{k,2}^\text{Comms-Tx}\big|^2}{\sigma_{w_\text{T}}^2}\right),
\label{eq:Rk2-max}
\earr
where  
$\big|\gain_{k,1}^\text{Comms-Rx}\big|^2 
= \big|\Upsilon_k(\widehat{\boldsymbol{\iota}}_{k \,|\, k-1})\big|^2$
and $\big|\gain_{k,2}^\text{Comms-Rx}\big|^2 
= \big|\Upsilon_k(\widehat{\boldsymbol{\iota}}_{k})\big|^2$
define the receive gains at the Comms Rx, with
$\Upsilon_k(\boldsymbol{\vartheta}) \eqdef \sum_{\ell=1}^L |c_{k,\ell}| \, e^{j \, [\bk(\boldsymbol{\iota}_k)-\bk(\boldsymbol{\vartheta})]^\trasp \, \pb_\ell}$
for $\boldsymbol{\vartheta} \in \{\widehat{\boldsymbol{\iota}}_{k \,|\, k-1},
\widehat{\boldsymbol{\iota}}_{k}\}$.

The optimization framework  developed here operates under the assumption 
that the predicted direction $\widehat{\boldsymbol{\iota}}_{k \,|\, k-1}$
is known  to the STAR-DCM at the beginning of the $k$-th time slot.
This assumption underpins all studies addressing the optimization
of the response of DCMs in wireless systems, where a control channel
between the optimizing entity and the DCMs is typically advocated. Hence, it is not a critical  assumption unique to STAR-DCMs.

\subsection{Entanglement and trade-off relationship}
\label{eq:ent-tradeoff}

A thorough evaluation of the overall communication performance requires
considering the sum rate $\rate_{k} = \rho_k \, \rate_{k,1}^\text{max} 
+ (1-\rho_k) \, \rate_{k,2}^\text{max}$,
where $\rho_k=I_{\text{ISAC},k} \, T_\text{u}/\Delta T \in (0,1]$
is the fraction of the $k$-th slot dedicated to ISAC and, consequently, $1-\rho_k=I_{\text{C},k} \, T_\text{u}/\Delta T$ is
the fraction of the slot duration exclusively allocated to communication.

The performance metrics $\rate_{k}$ and $\text{SNR}_{k,1}^\text{max}$
given by \eqref{eq:snr-max}
exhibit a complex interdepencence among the optimization variables: 
$\rho_k$ (i.e., time splitting between ISAC and communication-only sub-slots) 
and $\lambda_k$ (i.e., energy splitting between reflection and transmission), the predicted direction from the previous 
time slot (i.e., $\widehat{\boldsymbol{\iota}}_{k \,|\, k-1}$), 
the fine estimated direction from the current
time slot (i.e., $\widehat{\boldsymbol{\iota}}_{k}$), 
and the true direction (i.e., $\boldsymbol{\iota}_{k}$).
Specifically, $\widehat{\boldsymbol{\iota}}_{k \,|\, k-1}$
has been determined in the $(k-1)$-th slot and it is assumed to be given 
for optimization in the $k$-th slot. Meanwhile, 
$\widehat{\boldsymbol{\iota}}_{k}$ depends on both $\rho_k$ and $\lambda_k$:  
the sensing PCRB is inversely proportional to 
the time splitting variable $\rho_k$, determining 
the number of samples of the reflected signal processed
by the target state estimator, and to the energy splitting 
variable $\lambda_k$ related to $\text{SNR}_{k,1}^\text{max}$. Moreover, $\boldsymbol{\iota}_{k}$ is unknown at the ISAC
transceiver.

At this point, we are in the position to 
anticipate a fundamental trade-off between sensing and communication 
performance (see also Subsection~\ref{sec:solution}). When the error $\|\widehat{\boldsymbol{\iota}}_{k}
-\boldsymbol{\iota}_{k}\|_2$ is sufficiently small, 
it is apparent that $\rate_{k,2}^\text{ma} >\rate_{k,1}^\text{max}$. This is because the communication process in the
second sub-slot benefits from an accurate estimate
of the target direction  and   
$\big |\gamma_{\text{T},k,2}^{(\ell)} \big| 
> \big |\gamma_{\text{T},k,1}^{(\ell)} \big|$,
for each $\ell \in \{1,2, \ldots, L\}$. In this case,
$\rho_k$ is chosen as small as possible to
allocate a longer time duration to the communication-only sub-slot. 
However, an excessively small $\rho_k$ may lead to 
tracking failure, resulting in a poor estimate $\widehat{\boldsymbol{\iota}}_{k}$ 
of the true direction $\boldsymbol{\iota}_{k}$, which in its turn 
implies that the rate $\rate_{k,2}^\text{max}$ tends to zero.

\subsection{Optimization of the time and energy splitting variables}
\label{sec:solution}

To disentangle the intricate interplay among optimization variables and derive a practically realizable design that clearly enlightens the trade-off between 
time and energy splitting, a simple approach is followed, 
leveraging on the angular resolvability of the arrays and DCM.

Given that the STAR-DCM and the Comms Rx communicate 
within a short distance inside the mobile target, 
the approximation of $|c_{k,\ell}|$ with its mean value
$\mu_{c_k} \eqdef \Es[|c_{k,\ell}|]$ is reasonable. This mean value is assumed to remain relatively constant 
across DCM elements. Consequently, we might use the approximation
$\Upsilon_k(\boldsymbol{\vartheta}) \approx 
\mu_{c_k}  \sum_{\ell=1}^L 
e^{j \, [\bk(\boldsymbol{\iota}_k)-\bk(\boldsymbol{\vartheta})]^\trasp \, \pb_\ell}$
in \eqref{eq:Rk1-max} and \eqref{eq:Rk2-max}.

Let $\mathcal{A}_{k,1}$ denote the event when the predicted 
direction $\widehat{\boldsymbol{\iota}}_{k \,|\, k-1}$
is such that the errors $\left|\widehat{\theta}_{k \,|\, k-1} -\theta_k\right|$
and $\left|\widehat{\phi}_{k \,|\, k-1} -\phi_k\right|$
fall within the half-power beamwidth (HPBW) lobe \cite{Van-Trees} of the 
Comms Tx and Sens Rx (it inversely related to 
length of the array), as well as the HPBW lobe of the DCM (it is inversely related
to the horizontal and vertical lengths of the DCM). If $\mathcal{A}_{k,1}$ occurs, 
then the prediction error is below the resolution
capability of the arrays and DCM in the angular domain and, hence, we may assume 
that $\big|\gain_{k,1}^\text{Comms-Tx}\big|^2 \approx N$, 
$\big|\gain_{k,1}^\text{Sens-Rx}\big|^2 \approx M$, 
$\big|\gain_{k,1}^\text{DCM}\big|^2 \approx L^2$
in \eqref{eq:snr-max}, and 
$\big|\gain_{k,1}^\text{Comms-Rx}\big|^2 \approx \mu_{c_k}  \, L^2$ in 
\eqref{eq:Rk1-max}.
On the other hand if the complement $\mathcal{A}_{k,1}^\text{c}$ of the event 
$\mathcal{A}_{k,1}$ occurs, we might keep 
$\big|\gain_{k,1}^\text{Comms-Tx}\big|^2 \approx  
\big|\gain_{k,1}^\text{Sens-Rx}\big|^2 \approx  
\big|\gain_{k,1}^\text{DCM}\big|^2 \approx
\big|\gain_{k,1}^\text{Comms-Rx}\big|^2 \approx 0$.
As a consequence, by invoking the law of total expectation, 
the {\em average} value of the SNR in \eqref{eq:snr-max}
and the rate in \eqref{eq:Rk1-max} can be approximated as follows:
\barr
\overline{\text{SNR}}_{k,1}^\text{max}  \eqdef \Es[\text{SNR}_{k,1}^\text{max}]
& \approx \lambda_k \, \Prob[\mathcal{A}_{k,1}] \, \cost_{\text{R},k}, 
\label{eq:asnr}
\\
\overline{\rate}_{k,1}^\text{max}  \eqdef 
\Es[\rate_{k,1}^\text{max}] & \approx 
\Prob[\mathcal{A}_{k,1}] \, \log_2\Big(1+ (1-\lambda_k) \, \cost_{\text{T},k} \Big),
\label{eq:arate-1}
\earr
where the ensemble average is taken with respect to 
$\widehat{\boldsymbol{\iota}}_{k \,|\, k-1}$,
$\Prob[\cdot]$ denotes the probability of occurrence
of the event that appears inside the square brackets, 
and we have defined the constants 
$\cost_{\text{R},k} \eqdef \pot_{\text{R},k}\,
N \, M \, L^2/\sigma_{w_\text{R}}^2$ and 
$\cost_{\text{T},k} \eqdef \mu_{c_k} \, \pot_{\text{T},k}\,
N \, L^2/\sigma_{w_\text{T}}^2$.
It should be observed that $\Prob[\mathcal{A}_{k,1}]$ is independent
of $\rho_k$ and $\lambda_k$,
since $\widehat{\boldsymbol{\iota}}_{k \,|\, k-1}$
has been derived in the previous slot.

Applying the previous arguments to the second sub-slot, let
$\mathcal{A}_{k,2}$ denote the event when the estimated
direction $\widehat{\boldsymbol{\iota}}_{k}$
is such that the errors $\left|\widehat{\theta}_{k} -\theta_k\right|$
and $\left|\widehat{\phi}_{k} -\phi_k\right|$
fall within the HPBW lobe of the 
Comms Tx and DCM, given $\widehat{\boldsymbol{\iota}}_{k \,|\, k-1}$.
The rate \eqref{eq:Rk2-max} is approximated as 
\be
\overline{\rate}_{k,2}^\text{max}  \eqdef 
\Es[\rate_{k,2}^\text{max}] \approx 
\Prob[\mathcal{A}_{k,2}] \, \log_2\Big(1+\cost_{\text{T},k} \Big),
\label{eq:arate-2}
\ee
which provides the best guess at $\rate_{k,2}^\text{max}$ based on knowledge of
$\widehat{\boldsymbol{\iota}}_{k \,|\, k-1}$.
It should be noted that, according to
Fig.~\ref{fig:prediction_tracking}, 
besides depending on the value of the predicted direction 
$\widehat{\boldsymbol{\iota}}_{k \,|\, k-1}$ calculated
in the $(k-1)$-th slot,  the probability of $\mathcal{A}_{k,2}$ is 
also dictated by the tracking algorithm
used during the ISAC phase of the $k$-th slot. 
It can be shown (see, e.g., \cite{Meng.2024}) that $\Prob[\mathcal{A}_{k,2}] \to 1$ whenever 
the PCRB of the estimates 
$\widehat{\theta}_{k}$  
and $\widehat{\phi}_{k}$ becomes vanishingly  small, which
happens when the product 
$\rho_k \, \overline{\text{SNR}}_{k,1}^\text{max}$ gets large.

For each time slot $k$, the optimization model is formulated as
the constrained maximization of $\overline{\rate}_{k} = \rho_k \, \overline{\rate}_{k,1}^\text{max} 
+ (1-\rho_k) \, \overline{\rate}_{k,2}^\text{max}$ that, in light of the above 
remarks, simplifies as follows: 
\be
\max_{
\shortstack{\footnotesize $0 < \rho_k \le 1$  
\\ \footnotesize $0 < \lambda_k \le 1$}}  \, 
\rho_k \, \Prob[\mathcal{A}_{k,1}] \, 
\log_2\Big(1+ (1-\lambda_k) \, \cost_{\text{T},k} \Big) + 
(1-\rho_k) \, \Prob[\mathcal{A}_{k,2}] \, \log_2\Big(1+ \cost_{\text{T},k} \Big),
\label{eq:opt}
\ee
whose solution $(\rho_k^\text{opt},\lambda_k^\text{opt})$ 
can be obtained via a two-dimensional 
search over $\rho_k \in (0,1]$ and $\lambda_k \in (0,1]$.
Although problem \eqref{eq:opt} does not admit 
a closed-form solution, it suggests interesting insights regarding 
the trade-off between sensing and communication. 

\subsubsection{Poor prediction from the previous slot}
When $\Prob[\mathcal{A}_{k,1}] \to 0$, the first summand of the cost function
in \eqref{eq:opt} becomes negligible compared to the second one
and, thus, transmitting information to the Comms Rx during the 
first sub-slot is ineffective (in the information-theoretic sense). This
mandates that $\lambda_k^\text{opt}$ is close to $1$ as possible, i.e., the DCM
operates in a totally reflective state. The corresponding optimal value
of $\rho_k$ is mainly influenced by $\Prob[\mathcal{A}_{k,2}]$. If $\cost_{\text{R},k}$ is sufficiently high to offset the low value of $\Prob[\mathcal{A}_{k,1}]$ in \eqref{eq:asnr}, then $\rho_k^\text{opt}$
approaches zero, i.e., a short sensing interval is sufficient to 
provide an accurate estimate $\widehat{\boldsymbol{\iota}}_{k}$.
On the other hand, if $\cost_{\text{R},k}$ is such that 
$\overline{\text{SNR}}_{k,1}^\text{max} \approx \Prob[\mathcal{A}_{k,1}] \, \cost_{\text{R},k}  \ll 1$, then $\rho_k^\text{opt}$
increases to reduce the PCRB of the direction estimation 
and, consequently, the rate $\overline{\rate}_{k} \to 0$. 
 
\subsubsection{Accurate prediction from the previous slot}
When $\Prob[\mathcal{A}_{k,1}] \to 1$ and 
$\cost_{\text{R},k}$ takes on sufficiently large values, the 
duration of the ISAC sub-slot tends to last a negligible portion
of the $k$-th slot, i.e., $\rho_k^\text{opt} \to 0$.
In this case, the rate is mainly dictated
by the second sub-slot and $\lambda_k^\text{opt} \to 1$ in
order to increase $\Prob[\mathcal{A}_{k,2}]$. Conversely, when 
$\Prob[\mathcal{A}_{k,1}] \to 1$ and 
$\cost_{\text{R},k} \ll 1$, both summands 
of the cost function in \eqref{eq:opt} contribute to determine the
optimal values of $\rho_k$ and $\lambda_k$. In this latter case,
an increase in $\rho_k$ to cut down the sensing PCRB
is balanced by a reduction in $\lambda_k$ to maximize the rate
during the ISAC phase.

\section{Directions for future work}
\label{sec:future_directions}

The foundational principles of STAR-DCMs and ISAC as individual technologies are now well-established. Their integration has sparked significant interest only recently.
Opportunities for further development and exploration in this field include:

\begin{enumerate}

\item{\bf Trade-offs for Different STAR Mechanisms}: 
The effective use
of STAR-DCMs in ISAC systems relies on identifying mechanisms 
that balance performance,
power efficiency, robustness to channel distortions, and implementation cost. 
\item{\bf Synchronization Issues}: 
ISAC systems necessitate precise synchronization among Comms Tx, target, 
Comms Rx, and Sens Rx. Integrating DCMs (operating in STAR mode
or reflecting/transmitting-only mode) into ISAC networks 
introduces new synchronization challenges, 
particularly in 
bistatic and distributed deployments,
due to the limited signal 
processing capabilities of DCMs.
\item{\bf Temporal Modulation Schemes}:
   Advanced temporal modulation schemes \cite{Zhang:2018st}, such as periodic or almost-periodic changes within each time slot, could provide valuable time diversity that might be exploited for different ISAC network objectives.

\item{\bf Data-Driven Coding Schemes}:
   Overcoming the challenge of storing coding matrices for large metasurfaces in FPGA can be addressed by employing data-driven routines inspired by AI/ML techniques. These methods can ``learn'' the coding schemes by means of traditional approaches such as back-projection and nonlinear optimization methods \cite{Shan.2020}.

\item{\bf Higher Frequencies}:
   Terahertz (THz) signaling techniques (e.g., in $100-300$ GHz bands) provide a viable solution for achieving Tbit/s communication rates. Further development is needed for hardware implementations and signal-processing algorithms for ISAC systems using STAR-DCMs at these frequencies.

\item{\bf Security Considerations}:
   Integrating STAR-DCMs in ISAC systems introduces new hardware architectures and signal processing but also raises security concerns. Exploring potential physical-layer vulnerabilities and threats arising from metasurfaces is crucial to ensure the resilience of future ISAC systems \cite{Wei:2023me}.

\end{enumerate}

These areas indicate promising directions for significant advancements in the integration of STAR-DCMs into future ISAC systems.

\section{Conclusion}
\label{sec:summary}

This article has provided a tutorial on the fundamental signal-processing techniques that underpin the convergence between STAR-DCMs and ISAC. It has reviewed the EM design of STAR metasurfaces, with particular emphasis on practical hardware implementations of STAR-DCMs. The discussion has extended to ISAC protocols involving STAR-DCMs, highlighting key distinctions between communication and sensing links. Designs that effectively balance the demands of both sensing and communications have been discussed, taking into account the intricate interplay and synergy between such functionalities. Finally, potential pathways for advancing STAR-DCMs in future ISAC systems have been presented.

Overall, it is hoped that this overview will stimulate further cross-disciplinary exploration in this evolving paradigm.

\section{Acknowledgment}
\label{sec:ack}

The authors would like to thank the anonymous reviewers for
their stimulating comments and suggestions, which helped them
improve upon an earlier version of the manuscript.

The work of Francesco Verde was partially supported by the European Union under the Italian National Recovery and Resilience Plan (NRRP) of NextGenerationEU, partnership on ``Telecommunications of the Future" (PE00000001 - program ``RESTART").
The work of Vincenzo Galdi was partially supported by the University of Sannio via the FRA 2023 program.
The work of Lei Zhang and Tie Jun Cui was supported by the National Natural Science Foundation of China (62288101 and 62101123) and the 111 Project (111-2-05).

%
%
\bibliography{ISAC-STAR}

\begin{thebibliography}{10}
\providecommand{\url}[1]{#1}
\csname url@samestyle\endcsname
\providecommand{\newblock}{\relax}
\providecommand{\bibinfo}[2]{#2}
\providecommand{\BIBentrySTDinterwordspacing}{\spaceskip=0pt\relax}
\providecommand{\BIBentryALTinterwordstretchfactor}{4}
\providecommand{\BIBentryALTinterwordspacing}{\spaceskip=\fontdimen2\font plus
\BIBentryALTinterwordstretchfactor\fontdimen3\font minus
  \fontdimen4\font\relax}
\providecommand{\BIBforeignlanguage}[2]{{%
\expandafter\ifx\csname l@#1\endcsname\relax
\typeout{** WARNING: IEEEtran.bst: No hyphenation pattern has been}%
\typeout{** loaded for the language `#1'. Using the pattern for}%
\typeout{** the default language instead.}%
\else
\language=\csname l@#1\endcsname
\fi
#2}}
\providecommand{\BIBdecl}{\relax}
\BIBdecl

\bibitem{Stock.1948}
H.~Stockman, ``Communication by means of reflected power,'' \emph{Proc. IRE},
  vol.~36, no.~10, pp. 1196--1204, Oct. 1948.

\bibitem{Mealey.1963}
R.~M. Mealey, ``A method for calculating error probabilities in a radar
  communication system,'' \emph{IEEE Trans. Space Electron. Telem.}, vol.~9,
  no.~2, pp. 37--42, Jun. 1963.

\bibitem{Capolino}
F.~Capolino, \emph{Theory and Phenomena of Metamaterials}.\hskip 1em plus 0.5em
  minus 0.4em\relax Boca Raton: CRC Press, 2017.

\bibitem{DiRenzo.2022}
M.~Di~Renzo, F.~H. Danufane, and S.~Tretyakov, ``Communication models for
  reconfigurable intelligent surfaces: {From} surface electromagnetics to
  wireless networks optimization,'' \emph{{Proc. IEEE}}, vol. 110, no.~9, pp.
  1164--1209, Sep. 2022.

\bibitem{Nerini.2024}
M.~Nerini, S.~Shen, H.~Li, and B.~Clerckx, ``Beyond diagonal reconfigurable
  intelligent surfaces utilizing graph theory: {Modeling}, architecture design,
  and optimization,'' \emph{IEEE Trans. Wireless Commun.}, pp. 1--14, 2024,
  {Early Access}.

\bibitem{Cui:2014dm}
T.~J. Cui, M.~Q. Qi, X.~Wan, J.~Zhao, and Q.~Cheng, ``Coding metamaterials,
  digital metamaterials and programmable metamaterials,'' \emph{Light Sci.
  Appl.}, vol.~3, no.~10, p. e218, Oct. 2014.

\bibitem{YCui.2021}
Y.~Cui, F.~Liu, X.~Jing, and J.~Mu, ``Integrating sensing and communications
  for ubiquitous {IoT}: {Applications}, trends, and challenges,'' \emph{IEEE
  Netw.}, vol.~35, no.~5, pp. 158--167, Sep./Oct. 2021.

\bibitem{Esma.2022}
Z.~Esmaeilbeig, K.~V. Mishra, and M.~Soltanalian, ``{IRS}-aided radar:
  {Enhanced} target parameter estimation via intelligent reflecting surfaces,''
  in \emph{Proceedings of IEEE 12th IEEE Sens. Array Multichannel Signal
  Process. Workshop}, Trondheim, Norway, Jun. 2022, pp. 286--290.

\bibitem{Esma_Arxiv.2024}
\BIBentryALTinterwordspacing
T.~Esmaeilbeig, K.~V. Mishra, and M.~Soltanalian, ``Communication-assisted
  sensing in {6G} networks,'' 2024. [Online]. Available:
  \url{https://arxiv.org/abs/2402.14157}
\BIBentrySTDinterwordspacing

\bibitem{Zhang:2018tr}
L.~Zhang, R.~Y. Wu, G.~D. Bai, H.~T. Wu, Q.~Ma, X.~Q. Chen, and T.~J. Cui,
  ``Transmission-reflection-integrated multifunctional coding metasurface for
  full-space controls of electromagnetic waves,'' \emph{Adv. Funct. Mater.},
  vol.~28, no.~33, p. 1802205, Aug. 2018.

\bibitem{Bao:2021pr}
L.~Bao, Q.~Ma, R.~Y. Wu, X.~Fu, J.~Wu, and T.~J. Cui, ``Programmable
  reflection-transmission shared-aperture metasurface for real-time control of
  electromagnetic waves in full space,'' \emph{Adv. Sci.}, vol.~8, no.~15, p.
  2100149, Aug. 2021.

\bibitem{Mu.2022}
X.~Mu, Y.~Liu, L.~Guo, J.~Lin, and R.~Schober, ``Simultaneously transmitting
  and reflecting {(STAR) RIS} aided wireless communications,'' \emph{IEEE
  Trans. Wirel. Commun.}, vol.~21, no.~5, pp. 3083--3098, May 2022.

\bibitem{Xu.2022}
J.~Xu, Y.~Liu, X.~Mu, J.~T. Zhou, L.~Song, H.~V. Poor, and L.~Hanzo,
  ``Simultaneously transmitting and reflecting intelligent omni-surfaces:
  {Modeling} and implementation,'' \emph{{IEEE Veh. Technol. Mag.}}, vol.~17,
  no.~2, pp. 46--54, Sep./Oct. 2022.

\bibitem{Wang.2023}
Z.~Wang, X.~Mu, and Y.~Liu, ``{STARS} enabled integrated sensing and
  communications,'' \emph{IEEE Trans. Wireless Commun.}, vol.~22, no.~10, pp.
  6750--6765, Oct. 2023.

\bibitem{Liu.2023}
Z.~Liu, X.~Li, H.~Ji, H.~Zhang, and V.~C.~M. Leung, ``Toward
  {STAR-RIS}-empowered integrated sensing and communications: {Joint} active
  and passive beamforming design,'' \emph{IEEE Trans. Veh. Technol.}, vol.~72,
  no.~12, pp. 15\,991--16\,005, Dec. 2023.

\bibitem{Cai:2017he}
T.~Cai, G.~Wang, S.~Tang, H.~Xu, J.~Duan, H.~Guo, F.~Guan, S.~Sun, Q.~He, and
  L.~Zhou, ``High-efficiency and full-space manipulation of electromagnetic
  wave fronts with metasurfaces,'' \emph{Phys. Rev. Applied}, vol.~8, no.~3, p.
  034033, Sep. 2017.

\bibitem{Wang:2018sr}
X.~Wang, J.~Ding, B.~Zheng, S.~An, G.~Zhai, and H.~Zhang, ``Simultaneous
  realization of anomalous reflection and transmission at two frequencies using
  bi-functional metasurfaces,'' \emph{Sci. Rep.}, vol.~8, no.~1, p. 1876, Jan.
  2018.

\bibitem{Yue:2019da}
H.~Yue, L.~Chen, Y.~Yang, L.~He, and X.~Shi, ``Design and implementation of a
  dual frequency and bidirectional phase gradient metasurface for beam
  convergence,'' \emph{IEEE Antennas Wirel. Propag. Lett.}, vol.~18, no.~1, pp.
  54--58, Jan. 2019.

\bibitem{Liu:2020rs}
G.~Liu, H.~Liu, J.~Han, Y.~Mu, and L.~Li, ``Reconfigurable metasurface with
  polarization-independent manipulation for reflection and transmission
  wavefronts,'' \emph{J. Phys. D Appl. Phys.}, vol.~53, no.~4, p. 045107, Nov.
  2019.

\bibitem{Wu:2019dm}
R.~Y. Wu, L.~Zhang, L.~Bao, L.~W. Wu, Q.~Ma, G.~D. Bai, H.~T. Wu, and T.~J.
  Cui, ``Digital metasurface with phase code and reflection-transmission
  amplitude code for flexible full-space electromagnetic manipulations,''
  \emph{Adv. Opt. Mater.}, vol.~7, no.~8, p. 1801429, Apr. 2019.

\bibitem{Wang:2021ar}
H.~L. Wang, H.~F. Ma, M.~Chen, S.~Sun, and T.~J. Cui, ``A reconfigurable
  multifunctional metasurface for full-space controls of electromagnetic
  waves,'' \emph{Adv. Funct. Mater.}, vol.~31, no.~25, p. 2100275, Jun. 2021.

\bibitem{Zhang:2022io}
H.~Zhang, S.~Zeng, B.~Di, Y.~Tan, M.~Di~Renzo, M.~Debbah, Z.~Han, H.~V. Poor,
  and L.~Song, ``Intelligent omni-surfaces for full-dimensional wireless
  communications: Principles, technology, and implementation,'' \emph{IEEE
  Commun. Mag.}, vol.~60, no.~2, pp. 39--45, Feb. 2022.

\bibitem{Hu:2022ai}
Q.~Hu, J.~Zhao, K.~Chen, K.~Qu, W.~Yang, J.~Zhao, T.~Jiang, and Y.~Feng, ``An
  intelligent programmable omni-metasurface,'' \emph{Laser Photon. Rev.},
  vol.~16, no.~6, p. 2100718, Jun. 2022.

\bibitem{Wang:2022bh}
Y.~Wang, Y.~Ge, Z.~Chen, X.~Liu, J.~Pu, K.~Liu, H.~Chen, and Y.~Hao,
  ``Broadband high-efficiency ultrathin metasurfaces with simultaneous
  independent control of transmission and reflection amplitudes and phases,''
  \emph{IEEE Trans. Microwave Theory Techn.}, vol.~70, no.~1, pp. 254--263,
  Jan. 2022.

\bibitem{Yin:2023rt}
T.~Yin, J.~Ren, B.~Zhang, P.~Li, Y.~Luan, and Y.~Yin, ``Reconfigurable
  transmission-reflection-integrated coding metasurface for full-space
  electromagnetic wavefront manipulation,'' \emph{Adv. Opt. Mater.}, vol.~12,
  no.~2, p. 2301326, Jan. 2024.

\bibitem{Sun:2023tr}
S.~Sun, H.~F. Ma, Y.~T. Chen, and T.~J. Cui,
  ``Transmission-reflection-integrated metasurface with simultaneous amplitude
  and phase controls of circularly polarized waves in full space,'' \emph{Laser
  Photon. Rev.}, vol.~18, no.~3, p. 2300945, Mar. 2024.

\bibitem{Qin:2023tr}
Z.~Qin, Y.~Li, H.~Wang, C.~Li, C.~Liu, Z.~Zhu, Q.~Yuan, J.~Wang, and S.~Qu,
  ``Transmission reflection integrated programmable metasurface for real-time
  beam control and high efficiency transmission polarization conversion,''
  \emph{Ann. Phys.}, vol. 535, no.~1, p. 2200368, Jan. 2023.

\bibitem{Eghbali.2024}
Y.~Eghbali, S.~Faramarzi, S.~K. Taskou, M.~R. Mili, M.~Rasti, and E.~Hossain,
  ``Beamforming for {STAR-RIS}-aided integrated sensing and communication using
  meta {DRL},'' \emph{IEEE Wireless Commun. Lett.}, vol.~13, no.~4, pp.
  919--923, Apr. 2024.

\bibitem{Meng.2024}
K.~Meng, Q.~Wu, W.~Chen, and D.~Li, ``Sensing-assisted communication in
  vehicular networks with intelligent surface,'' \emph{IEEE Trans. Veh.
  Technol.}, vol.~73, no.~1, pp. 876--893, Jan. 2024.

\bibitem{Li.2024}
M.~Li, S.~Zhang, Y.~Ge, Z.~Li, F.~Gao, and P.~Fan, ``{STAR-RIS} aided
  integrated sensing and communication over high mobility scenario,''
  \emph{IEEE Trans. Commun.}, pp. 1--15, 2024, {Early Access}.

\bibitem{Sun.2024}
G.~Sun, Y.~Zhang, W.~Hao, Z.~Zhu, X.~Li, and Z.~Chu, ``Joint beamforming
  optimization for {STAR-RIS} aided {NOMA ISAC} systems,'' \emph{IEEE Wireless
  Commun. Lett.}, vol.~13, no.~4, pp. 1009--1013, Apr. 2024.

\bibitem{Wang.2024}
Y.~Wang, Z.~Yang, J.~Cui, P.~Xu, G.~Chen, T.~Q.~S. Quek, and R.~Tafazolli,
  ``Optimizing the fairness of {STAR-RIS} and {NOMA} assisted integrated
  sensing and communication systems,'' \emph{IEEE Trans. Wireless Commun.},
  vol.~23, no.~6, pp. 5895--5907, Jun. 2024.

\bibitem{Xue_2024}
N.~Xue, X.~Mu, Y.~Liu, and Y.~Chen, ``{NOMA} assisted full space
  {STAR-RIS-ISAC},'' \emph{IEEE Trans. Wireless Commun.}, pp. 1--15, 2024,
  {Early Access}.

\bibitem{Wei.2024}
W.~Wei, X.~Pang, C.~Xing, N.~Zhao, and D.~Niyato, ``{STAR-RIS} aided secure
  {NOMA} integrated sensing and communication,'' \emph{IEEE Trans. Wireless
  Commun.}, pp. 1--14, 2024, {Early Access}.

\bibitem{Weijie.2021}
W.~Yuan, F.~Liu, C.~Masouros, J.~Yuan, D.~W.~K. Ng, and N.~González-Prelcic,
  ``Bayesian predictive beamforming for vehicular networks: {A} low-overhead
  joint radar-communication approach,'' \emph{{IEEE Trans. Wireless Commun.}},
  vol.~20, no.~3, pp. 1442--1456, Mar. 2021.

\bibitem{Shab.2022}
J.~Shabanpour and C.~R. Simovski, ``Angular and polarization stability of
  broadband reconfigurable intelligent surfaces of binary type,'' \emph{IEEE
  Access}, vol.~10, pp. 126\,253--126\,268, 2022.

\bibitem{Li.2023}
Q.~Li, P.~Sisk, A.~Kannan, T.~Yoo, T.~Luo, G.~Shah, B.~Manjunath,
  C.~Samarathungage, M.~T. Boroujeni, H.~Pezeshki, and H.~Joshi, ``Machine
  learning based time domain millimeter-wave beam prediction for {5G}-advanced
  and beyond: {Design}, analysis, and over-the-air experiments,'' \emph{IEEE J.
  Select. Areas Commun.}, vol.~41, no.~6, pp. 1787--1809, Jun. 2023.

\bibitem{Van-Trees}
{H.L. Van Trees}, \emph{{Optimum Array Processing: Part IV of Detection,
  Estimation, and Modulation Theory}}.\hskip 1em plus 0.5em minus 0.4em\relax
  Hoboken, NJ, USA: Wiley, 2004.

\bibitem{Zhang:2018st}
L.~Zhang, X.~Q. Chen, S.~Liu, Q.~Zhang, J.~Zhao, J.~Y. Dai, G.~D. Bai, X.~Wan,
  Q.~Cheng, G.~Castaldi, V.~Galdi, and T.~J. Cui, ``Space-time-coding digital
  metasurfaces,'' \emph{Nat. Commun.}, vol.~9, p. 4334, Oct. 2018.

\bibitem{Shan.2020}
T.~Shan, X.~Pan, M.~Li, S.~Xu, and F.~Yang, ``Coding programmable metasurfaces
  based on deep learning techniques,'' \emph{IEEE J. Emerg. Sel. Top. Circuits
  Syst.}, vol.~10, no.~1, pp. 114--125, Mar. 2020.

\bibitem{Wei:2023me}
M.~Wei, H.~Zhao, V.~Galdi, L.~Li, and T.~J. Cui, ``Metasurface-enabled smart
  wireless attacks at the physical layer,'' \emph{Nat. Electron.}, vol.~6,
  no.~8, pp. 610--618, Aug. 2023.

\end{thebibliography}
\bibliographystyle{IEEEtran}

\end{document}